%% file: paper3.tex
\documentclass[a4paper,fleqn,usenatbib]{mnras}

\usepackage{newtxtext}
\usepackage{newtxmath}

\usepackage[T1]{fontenc}
\usepackage{ae,aecompl}

\usepackage{graphicx}	%

\ifdefined\Bbbk{}
    \let\Bbbk\relax
\else
\fi
\ifdefined\bigtimes{}
    \let\bigtimes\relax
\else
\fi
\usepackage[separate-uncertainty=true,
            multi-part-units=single]{siunitx}
\usepackage{pgf}
\usepackage{xprintlen}
\usepackage{booktabs}

\usepackage{mathabx}  %

\usepackage{wasysym}

\usepackage{sparklines}  %
\usepackage{orcidlink} %
\usepackage{lipsum}  %

\input{star-commands.tex}

\hypersetup{pdfauthor={D. Evensberget et al.}}  %

\newcommand{\batsrus}{{\textsc{bats-r-us}}}
\newcommand{\swmf}{{\textsc{swmf}}}
\newcommand{\awsom}{{\textsc{awsom}}}
\newcommand{\toupies}{TOUPIES}

\newcommand{\SF}{{\Phi}}  %

\definecolor{darkgreen}{rgb}{0.0, 0.5, 0.0}

\newcommand{\Wind}{\textsc{w}} %
\newcommand{\Rot}{\text{rot}} %
\newcommand{\Alfven}{{\textsc{a}}} %
\newcommand{\Mag}{\text{mag}} %
\newcommand{\Open}{\text{open}} %
\newcommand{\Star}{\text{\bigstar}} %
\newcommand{\CS}{{{\scriptscriptstyle B}}} %

\newcommand{\ZDI}{{\mathchoice{}{}{\scriptscriptstyle}{}\mathrm{ZDI}}}

\renewcommand{\Earth}{{\mathchoice{}{}{\scriptscriptstyle}{}\oplus}} %
\renewcommand{\Sun}{{\mathchoice{}{}{\scriptscriptstyle}{}\odot}} %
\renewcommand{\Star}{{\mathchoice{}{}{\scriptscriptstyle}{}\bigstar}} %

\defcitealias{2016MNRAS.457..580F}{F16}
\defcitealias{2018MNRAS.474.4956F}{F18}

\defcitealias{paper1}{Paper I}
\defcitealias{paper2}{Paper II}

\DeclareSIUnit\year{yr}
\DeclareSIUnit\astronomicalunit{au}
\DeclareSIUnit\parsec{pc}
\DeclareSIUnit\erg{erg}
\DeclareSIUnit\gauss{G}
\DeclareSIUnit\mSun{\mbox{\(M_\Sun\)}}
\DeclareSIUnit\rSun{\mbox{\(R_\Sun\)}}
\DeclareSIUnit\mEarth{\mbox{\(M_\Earth\)}}
\DeclareSIUnit\rEarth{\mbox{\(R_\Earth\)}}

\renewcommand{\vec}[1]{\boldsymbol{#1}} %
\newcommand{\uvec}[1]{\boldsymbol{\hat{#1}}} %

\newcommand\thefont{\expandafter\string\the\font} %

\renewcommand{\propto}{\mathrel{\mathchar"939}} %
\newcommand{\mysim}{{\sim}} %

\newcommand{\appropto}{\mathrel{\vcenter{
  \offinterlineskip\halign{\hfil$##$\cr  %
    \propto\cr\noalign{\kern2pt}\sim\cr\noalign{\kern-2pt}}}}}  %

\title[Winds of the Pleiades, AB Doradus, Columba and \(\beta\) Pictoris]{The winds of young Solar-type stars in the Pleiades, AB Doradus, Columba and \(\beta\) Pictoris}
\author[D. Evensberget et al.]{D. Evensberget \orcidlink{0000-0001-7810-8028}\(^{1,2}\)\thanks{E-mail: \href{mailto:evensberget@strw.leidenuniv.nl}{evensberget@strw.leidenuniv.nl}},
    S. C. Marsden  \orcidlink{0000-0001-5522-8887}\(^{2}\),
    B. D. Carter   \orcidlink{0000-0003-0035-8769}\(^{2}\),
    R. Salmeron    \orcidlink{0000-0002-1956-4493}\(^{2}\),
    A. A. Vidotto  \orcidlink{0000-0001-5371-2675}\(^{1}\),
    \newauthor
    C. P. Folsom    \orcidlink{0000-0002-9023-7890}\(^{3,4}\),
    R. D. Kavanagh  \orcidlink{0000-0002-1486-7188}\(^{5,1}\),
    J. S. Pineda    \orcidlink{0000-0002-4489-0135}\(^{6}\),
    F. A. Driessen  \orcidlink{0000-0003-3005-7377}\(^{1}\),
    and
    K. M. Strickert  \orcidlink{0000-0001-6584-5969}\(^{1}\)
    \\
    \(^{1}\)Leiden Observatory, Leiden University, PO Box 9513, 2300 RA Leiden, The Netherlands
    \\
    \(^{2}\)Centre for Astrophysics, University of Southern Queensland, Toowoomba, Queensland 4350, Australia
    \\
    \(^{3}\)Tartu Observatory, University of Tartu, Observatooriumi 1, T\~{o}ravere, 61602 Tartumaa, Estonia
    \\
    \(^{4}\)University of Western Ontario, Department of Physics \& Astronomy, London, ON, N6A 3K7, Canada
    \\
    \(^{5}\)ASTRON, The Netherlands Institute for Radio Astronomy, Oude Hogeveensedijk 4, 7991PD, Dwingeloo, The Netherlands
    \\
    \(^{6}\)University of Colorado Boulder, Laboratory for Atmospheric and Space Physics, 3665 Discovery Drive, Boulder, CO 80303, USA 
}

\date{Accepted XXX.\@ Received YYY;\@ in original form ZZZ}

\pubyear{2023}
  
\begin{document}
\label{firstpage} %
\pagerange{\pageref{firstpage}--\pageref{lastpage}}
\maketitle

\begin{abstract}
    \input{abstract}
\end{abstract}


\input{paper3-body}

\bibliographystyle{mnras}
\bibliography{bibliography} %

\clearpage
\appendix
\input{reprocessed-table.tex}
\input{pooled-series.tex}

\bsp	%
\label{lastpage} %
\end{document}

%% file: star-commands.tex
\usepackage{tikz}
\usetikzlibrary{shapes.geometric}
\usepackage{xstring}

\definecolor{C0}{HTML}{1f77b4}  %
\definecolor{C1}{HTML}{ff7f0e}
\definecolor{C2}{HTML}{2ca02c}
\definecolor{C3}{HTML}{d62728}
\definecolor{C4}{HTML}{9467bd}
\definecolor{C5}{HTML}{8c564b}
\definecolor{C6}{HTML}{e377c2}
\definecolor{C7}{HTML}{7f7f7f}

\DeclareRobustCommand{\C}[1]{\SimCase{#1}}  

\newcommand{\SimCase}[2]{%
    \IfStrEqCase{#2}{
        {Z}{\NumScaled{#1}{#2}{black}{Sun}}%
        {0}{\NumScaled{#1}{#2}{C0}{Mel25\nobreakdash-5}}%
        {1}{\NumScaled{#1}{#2}{C0}{Mel25\nobreakdash-21}}%
        {2}{\NumScaled{#1}{#2}{C0}{Mel25\nobreakdash-43}}%
        {3}{\NumScaled{#1}{#2}{C0}{Mel25\nobreakdash-151}}%
        {4}{\NumScaled{#1}{#2}{C0}{Mel25\nobreakdash-179}}%
        {5}{\NumScaled{#1}{#2}{C1}{AV\,523}}%
        {6}{\NumScaled{#1}{#2}{C1}{AV\,1693}}%
        {7}{\NumScaled{#1}{#2}{C1}{AV\,1826}}%
        {8}{\NumScaled{#1}{#2}{C1}{AV\,2177}}%
        {9}{\NumScaled{#1}{#2}{C1}{TYC\,1987}}%
        {A}{\NumScaled{#1}{#2}{C2}{DX\,Leo}}%
        {B}{\NumScaled{#1}{#2}{C2}{EP\,Eri}}%
        {C}{\NumScaled{#1}{#2}{C2}{HH\,Leo}}%
        {D}{\NumScaled{#1}{#2}{C2}{V439\,And}}%
        {E}{\NumScaled{#1}{#2}{C2}{V447\,Lac}}%
        {F}{\NumScaled{#1}{#2}{C3}{HII\,296}}%
        {G}{\NumScaled{#1}{#2}{C3}{HII\,739}}%
        {H}{\NumScaled{#1}{#2}{C3}{PELS\,031}}%
        {I}{\NumScaled{#1}{#2}{C4}{BD\nobreakdash-07\,2388}}%
        {J}{\NumScaled{#1}{#2}{C4}{HD\,6569}}%
        {K}{\NumScaled{#1}{#2}{C4}{HIP\,10272}}%
        {L}{\NumScaled{#1}{#2}{C4}{HIP\,76768}}%
        {M}{\NumScaled{#1}{#2}{C4}{LO\,Peg}}%
        {N}{\NumScaled{#1}{#2}{C4}{PW\,And}}%
        {O}{\NumScaled{#1}{#2}{C4}{TYC\,0486}}%
        {P}{\NumScaled{#1}{#2}{C4}{TYC\,5164}}%
        {Q}{\NumScaled{#1}{#2}{C5}{BD-16\,351}}%
        {R}{\NumScaled{#1}{#2}{C6}{HIP\,12545}}%
        {S}{\NumScaled{#1}{#2}{C6}{TYC\,6349}}%
        {T}{\NumScaled{#1}{#2}{C6}{TYC\,6878}}%
    }
    [(STAR NOT FOUND)]%
}

\newcommand{\SimSymbol}[2]{%
    \IfStrEqCase{#2}{
        {Z}{\NumScaled{#1}{#2}{black}}%
        {0}{\NumScaled{#1}{#2}{C0}}
        {1}{\NumScaled{#1}{#2}{C0}}
        {2}{\NumScaled{#1}{#2}{C0}}
        {3}{\NumScaled{#1}{#2}{C0}}
        {4}{\NumScaled{#1}{#2}{C0}}
        {5}{\NumScaled{#1}{#2}{C1}}
        {6}{\NumScaled{#1}{#2}{C1}}
        {7}{\NumScaled{#1}{#2}{C1}}
        {8}{\NumScaled{#1}{#2}{C1}}
        {9}{\NumScaled{#1}{#2}{C1}}
        {A}{\NumScaled{#1}{#2}{C2}}
        {B}{\NumScaled{#1}{#2}{C2}}
        {C}{\NumScaled{#1}{#2}{C2}}
        {D}{\NumScaled{#1}{#2}{C2}}
        {E}{\NumScaled{#1}{#2}{C2}}
        {F}{\NumScaled{#1}{#2}{C3}}
        {G}{\NumScaled{#1}{#2}{C3}}
        {H}{\NumScaled{#1}{#2}{C3}}
        {I}{\NumScaled{#1}{#2}{C4}}
        {J}{\NumScaled{#1}{#2}{C4}}
        {K}{\NumScaled{#1}{#2}{C4}}
        {L}{\NumScaled{#1}{#2}{C4}}
        {M}{\NumScaled{#1}{#2}{C4}}
        {N}{\NumScaled{#1}{#2}{C4}}
        {O}{\NumScaled{#1}{#2}{C4}}
        {P}{\NumScaled{#1}{#2}{C4}}
        {Q}{\NumScaled{#1}{#2}{C5}}
        {R}{\NumScaled{#1}{#2}{C6}}
        {S}{\NumScaled{#1}{#2}{C6}}
        {T}{\NumScaled{#1}{#2}{C6}}
    }
    [(STAR NOT FOUND)]%
}

\newcommand{\NumScaled}[3]{%
    \IfStrEqCase{#1}{
        {1}{\Unscaled{#2}{#3}}
        {2}{\TwoScaled{#2}{#3}\(2\times\)}
        {5}{\Scaled{#2}{#3}\(5\times\)}
    }
}

\newcommand{\Unscaled}[2] %
{\ignorespaces  %
    \tikz[baseline={([yshift=-.75ex]current bounding box.center)}]
    {
        \node
        [
            circle,
            anchor=base,
            text=white,
            fill=#2,
            inner sep=0.7pt,
            minimum width=0.24cm
        ]
        (M)
        {
            \tiny{#1}
        };
    }\,\ignorespaces  %
}
\newcommand{\TwoScaled}[2] %
{\ignorespaces  %
    \tikz[baseline={([yshift=-.75ex]current bounding box.center)}]
    {
        \node
        [
            star,
            anchor=base,
            star points=4,
            shape border rotate=45,
            star point ratio=2,
            text=white,
            fill=#2,
            inner sep=0.0pt,
            minimum width=0.37cm
        ](M)  %
        {
            \tiny{#1}
        };
    }\,\ignorespaces  %
}
\newcommand{\Scaled}[2] %
{\ignorespaces  %
    \tikz[baseline={([yshift=-.75ex]current bounding box.center)}]
    {
        \node
        [
            star,
            anchor=base,
            star point ratio=2,
            text=white,
            fill=#2,
            inner sep=0.0pt,
            minimum width=0.37cm
        ](M)  %
        {
            \tiny{#1}
        };
    }\,\ignorespaces  %
}

%% file: abstract.tex
    Solar-type stars, which shed angular momentum via magnetised stellar winds, enter the main sequence with a wide range of rotational periods \(P_\Rot\). 
    This initially wide range of rotational periods contracts and has mostly vanished by a stellar age \(t\sim\SI{0.6}{\giga\year}\), after which Solar-type stars spin according to the Skumanich relation \(P_\Rot\propto\sqrt t\). Magnetohydrodynamic stellar wind models can improve our understanding of this convergence of rotation periods.
    We present wind models of fifteen young Solar-type stars aged from \SI{\sim24}{\mega\year} to 
    \SI{\sim0.13}{\giga\year}. 
    With our previous wind models of stars aged \SI{\sim0.26}{\giga\year} and \SI{\sim0.6}{\giga\year} we obtain thirty consistent three-dimensional wind models of stars mapped with Zeeman-Doppler imaging---the largest such set to date. The models provide good cover of the pre-Skumanich phase of stellar spin-down in terms of rotation, magnetic field, and age. 
    We find that the mass loss rate
    \(\dot M\propto\Phi^{\num{0.9\pm0.1}}\) with a residual spread of \SI{\sim150}{\percent}
    and that the wind angular momentum loss rate
    \(\dot J\propto{}P_\Rot^{-1} \Phi^{1.3\pm0.2}\) 
    with a residual spread of \SI{\sim500}{\percent} where \(\Phi\) is the unsigned surface magnetic flux. 
    When comparing different magnetic field scalings for each single star we find a gradual reduction in the power-law exponent with increasing magnetic field strength. 

%% file: paper3-body.tex
\section{Introduction}
\begin{table*}
    \centering
    \sisetup{
        table-figures-decimal=2,
        table-figures-integer=2,
        table-figures-uncertainty=2,
        table-figures-exponent = 0,
        table-number-alignment=center,
        round-mode=places,
        round-precision=1
    }
    \caption{
        Observed stellar quantities from~\citetalias{2016MNRAS.457..580F,2018MNRAS.474.4956F} and references therein. 
        In this work we denote the modelled stars by the abbreviated names given in the case name column. Each entry in the case column is prepended by a coloured symbol which may be used to identify each stellar case in the plots throughout this paper. The full name column refers to the name used in~\citetalias{2016MNRAS.457..580F,2018MNRAS.474.4956F}. The assoc.\ column gives the stellar association to which the stars belong. The stellar type and effective temperature are given in the type and \(T_\text{eff}\) columns respectively. The period of rotation is given in the \(P_\Rot\) column, age is given in the eponymous column, and the radius and mass are given in the \(R\) and \(M\) columns respectively, in terms of the Solar radius \(R_\Sun\) and Solar mass \(M_\Sun\). The reference from which the data is taken is given in the ref.\ column. When not separately available in the reference literature, the spectral type is calculated from \(T_\text{eff}\) following~\citet{2013ApJS..208....9P}; this is denoted by a dagger (\(\dag\)) symbol in the type column.
    }\label{tab:observed_quantities}
    \input{tables/table1.tex}
\end{table*}
The pre-Skumanich spin-down phase of a cool star's life is the era between the end of the contractive spin-up phase and the~\citet{1972ApJ...171..565S} spin-down phase where the stellar period of rotation \(P_\Rot \propto \sqrt t\)\,, where \(t\) is the stellar age.  For Solar-type stars the pre-Skumanich spin-down phase ends around \SI{0.6}{\giga\year}~\citep{2013A&A...556A..36G,2015A&A...577A..98G}.
As stars enter the pre-Skumanich spin-down phase with a spread of rotational periods of at least an order of magnitude~\citep{1993AJ....106..372E}, angular momentum shedding in the pre-Skumanich spin-down phase must be instrumental in permitting the convergence of rotational periods with increasing stellar age.

Angular momentum shedding occurs by means of magnetised stellar winds~\citep{1962AnAp...25...18S}. 
Semi-empirical models of angular momentum loss on the general form 
\(\dot J \propto \dot M P_\Rot^{-1} R_\Alfven^n\) 
have been proposed in many studies~\citep{1967ApJ...148..217W,1968MNRAS.138..359M,1984LNP...193...49M,1988ApJ...333..236K}. In these models \(\dot M\) is the stellar winds mass loss rate and \(R_\Alfven\) denotes the average radius of the Alfvén surface where the wind's kinetic energy and magnetic energy have similar magnitudes. 
The average Alfvén radius \(R_\Alfven\) depends on the large-scale coronal magnetic field strength \(|B|\) but also on \(\dot M\) as both the wind kinetic energy and mass loss are determined by the wind speed and density.
The wind mass loss rate \(\dot M\) is poorly constrained by observations, particularly for stars younger than \SI{0.6}{\giga\year}~\citep{2005ApJ...628L.143W,2014ApJ...781L..33W}; this uncertainty propagates to our knowledge of \(\dot J\) and \(R_\Alfven\). The exponent \(n \lesssim 2\) is a geometric parameter which decreases with increasing complexity of the large-scale coronal magnetic field. 

The magnitude of the large-scale photospheric magnetic field decreases with increasing \(P_\Rot\)~\citep{2014MNRAS.444.3517M}, as does stellar magnetic activity in general~\citep{1984ApJ...279..763N}.  This decrease is presumably caused by a throttling of the internal stellar dynamo as the amount of available rotational energy decreases.
Qualitative dynamo changes with increasing \(P_\Rot\)~\citep[see][]{2003ApJ...586..464B,2011MNRAS.418L.133M,2011MNRAS.411.1301J,2011MNRAS.413.1922M,2011MNRAS.413.1939M,2014ApJ...789..101B} would also affect the complexity of the photospheric magnetic field as it emerges from lower stellar layers. 

Both the strength and complexity of the large-scale stellar photospheric magnetic field can be studied with spectropolarimetric stellar observations and Zeeman-Doppler imaging~\citep[ZDI,][]{1989A&A...225..456S,1997MNRAS.291..658D}. 
While the average large-scale magnetic field intensity found using Zeeman-Doppler imaging is subject to uncertainty~\citep{2019MNRAS.483.5246L},
on average the ZDI-derived field decays as \(|B_\ZDI|\propto P_\Rot^{\num{-1.32\pm0.14}}\), however, the relation is subject to significant scatter~\citep{2014MNRAS.441.2361V}. 
Zeeman-Doppler imaging  tends to yield more complex magnetic field geometries in younger stars~\citep{2018MNRAS.474.4956F}, but this may be due to the technique's increased resolving power for rapidly spinning stars~\citep{2010MNRAS.407.2269M}.

The early three-dimensional numerical wind models by~\citet{2009ApJ...699..441V,2012MNRAS.423.3285V,2010ApJ...721...80C} show that field complexity can significantly affect the angular momentum loss rate \(\dot J\). This highlights the need for incorporating realistic large-scale magnetic fields when studying the \(|B_\ZDI|\)-\(P_\Rot\) relation. Numerical models also suggest that \(\dot M\) (and thus also \(\dot J\)) may be inhibited by a complex magnetic field~\citep{2015ApJ...807L...6G}.

In this work we investigate the effect of the photospheric magnetic field and  stellar rotation rate on wind mass- and angular momentum loss rates as well as other parameters of interest by driving the state of the art numerical magnetohydrodynamic Alfvén wave solar model 
\citep[\awsom{},][]{2013ApJ...764...23S, 2014ApJ...782...81V} with the photospheric radial magnetic field from the magnetic maps of young, Solar-type stars published by~\citet[][hereafter~\citetalias{2016MNRAS.457..580F,2018MNRAS.474.4956F}]{2016MNRAS.457..580F,2018MNRAS.474.4956F}. We consider eleven stars with well-constrained ages in the \SI{\sim 130}{\mega\year} old Pleaides and AB~Doradus clusters and four very young stars in the 
\SI{\sim 42}{\mega\year} old Columba association and \SI{\sim 24}{\mega\year} old \(\beta\)~Pictoris association. 

By letting the resulting three-dimensional stellar wind models relax towards a steady-state, we obtain simultaneous simulated values for the wind mass loss rate \(\dot M\), wind angular momentum loss rate \(\dot J\) and the Alfvén radius \(R_\Alfven\), wind pressure for an Earth-like planet\footnote{By `Earth-like planet' we restrict ourselves to mean a planet whose radius, orbit and magnetic field are similar to those of the present-day Earth.}, as well as many other quantities of interest.  This permits us to investigate the degree to which the coronal magnetic field complexity affects stellar spin-down.

For Solar-type stars the pre-Skumanich spin-down phase ends around \SI{0.6}{\giga\year}~\citep{2013A&A...556A..36G,2015A&A...577A..98G}. By combining the results of this work with our previous wind models of \SI{\sim0.6}{\giga\year} old Solar-type stars in the 
Hyades~\citep[][hereafter~\citetalias{paper1}]{paper1}
and
\SI{\sim0.26}{\giga\year} old stars in  
Coma Berenices 
and Hercules-Lyra~\citep[][hereafter~\citetalias{paper2}]{paper2}, we get good age coverage of the pre-Skumanich spin-down period and a large sample of thirty stellar models from which trends may be robustly extracted.

This paper is outlined as follows: 
In Section~\ref{sec:Observations} we briefly describe the spectropolarimetric observations upon which the magnetic maps used in this work are based, as well as the process by which the surface magnetic field maps are obtained.
In Section~\ref{sec:Simulations}  we describe the stellar wind model in terms of physical effects included, model equations, three-dimensional numerical model, and boundary conditions.
In Section~\ref{sec:results} we present visualisations of the wind model coronal magnetic field, Alfvén surface and current sheet, and wind pressure out to \SI{1}{\astronomicalunit}, and tabulate aggregate wind parameters calculated from the three-dimensional wind maps.
In Section~\ref{sec:Discussion}   we investigate the effects of scaling the magnetic field, the statistical trends and variation of key wind parameters when plotted against the magnetic field strength and related parameters.
In Section~\ref{sec:Conclusions}  we conclude our work and put our results into context with other recent works in the field.

\section{Observations}\label{sec:Observations}
\begin{figure*}
    \centering
    \includegraphics{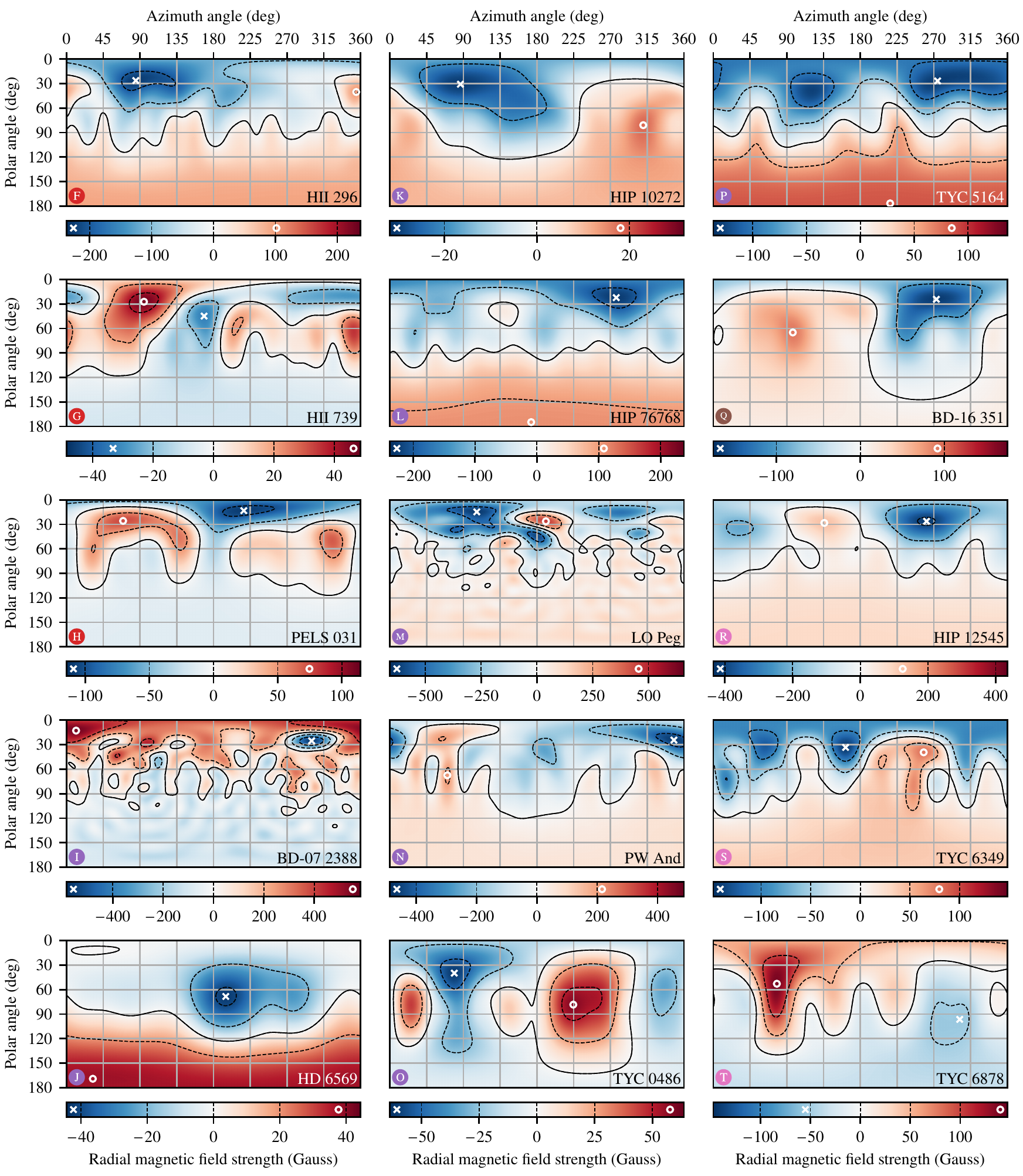}
    \caption{
        Surface magnetic radial field geometry~\citepalias{2016MNRAS.457..580F,2018MNRAS.474.4956F} and field strength in Gauss of the stars modelled in this study. The case name from Table~\ref{tab:observed_quantities} of each star and an associated identifying symbol is shown inside the bottom edge of each plot. The polar angle is zero at the stellar pole of rotation, and the azimuth angle is measured around the stellar rotational equator. The continuous black line represents the \(B_r=0\) contour and the dashed lines represents increments as indicated on the associated colour scale below each plot. The cross and circle symbols indicate the smallest and largest value of \(B_r\) respectively. The extremal values of \(B_r\) are also indicated in the colour bar. As \(\ell_\text{max} = 15\) in the ZDI maps, the minimum resolution in these plots is \SI{\sim 12}{\degree}.
    }\label{fig:magnetograms-zdi}
\end{figure*}

The stellar surface magnetic maps used in this work were derived from Zeeman-Doppler Imaging~\citep{1989A&A...225..456S} observations and modelling by~\citetalias{2016MNRAS.457..580F,2018MNRAS.474.4956F}. As part of `TOwards Understanding the sPIn Evolution of Stars' (\toupies{})\footnote{See the TOUPIES project page at \url{http://ipag.osug.fr/Anr_Toupies/}.} project, spectropolarimetric observations in circularly polarised and unpolarised light 
of the stellar targets in Table~\ref{tab:observed_quantities} 
were carried out using the ESPaDOnS instrument~\citep{2003ASPC..307...41D,2012MNRAS.426.1003S} at the Canada-Hawaii-France Telescope (CHFT) and using the Narval instrument~\citep{2003EAS.....9..105A} at the Télescope Bernard Lyot (TBL) between 2009 and 2015.
The ESPaDOnS observations of 
{BD\nobreakdash-07\,2388} and {HD\,6569} 
were part of the 
\emph{History of the Magnetic Sun}\footnote{Available at \url{https://www.cfht.hawaii.edu/Science/HMS/}.}
Large Program at the CFHT.\@
The \toupies{} reduced spectra can be downloaded from the Polarbase~\citep{1997MNRAS.291..658D,2014PASP..126..469P} website\footnote{Polarbase is available at \url{http://polarbase.irap.omp.eu/}.}.
Each star was observed over a period of a few weeks to obtain good phase coverage of the star while minimising intrinsic stellar variation. The resulting surface magnetic field maps are published in~\citepalias{2016MNRAS.457..580F,2018MNRAS.474.4956F}\footnote{
    The Zeeman-Doppler imaging software of \citeauthor{2016MNRAS.457..580F} is  available at \url{https://github.com/folsomcp/ZDIpy} as a Python package.
}; the radial component of the maps, which are used to drive the wind models of this work, are reproduced in Fig.~\ref{fig:magnetograms-zdi}.

\subsection{Magnetic mapping with Zeeman-Doppler imaging}\label{sec:mag-map-zdi}
This section gives a short overview of the Zeeman-Doppler imaging technique on which the stellar magnetic maps are based. 
Zeeman-Doppler imaging~\citep[ZDI,][]{1989A&A...225..456S} is a tomographic technique that permits the reconstruction of the stellar surface magnetic field. The technique requires high-resolution spectropolarimetric observations from multiple epochs in order to reconstruct a two-dimensional image of the vector surface magnetic field.
In ZDI, least-squares deconvolution~\citep[LSD,][]{1997A&A...326.1135D,2010A&A...524A...5K} is used to combine magnetically sensitive spectral line profiles in velocity space into a single profile with a much higher signal-to-noise ratio than individual lines.

By comparing the resulting circularly polarised and unpolarised LSD line profiles, and by combining multiple LSD profiles from different time and stellar phases, it is possible to reconstruct a two-dimensional surface map of the large-scale magnetic features on the stellar surface in a robust way~\citep{2000MNRAS.318..961H}. Linearly polarised light may also be included and has the potential to break degeneracies, but as the linearly polarised signal is \(\mysim10\) times weaker than the circularly polarised signal, this is only feasible for a subset of the stars to which ZDI has been applied, see~\citet{2015ApJ...805..169R}. The review by~\citet{2009ARA&A..47..333D} gives an overview of ZDI applied to cool stars.

The amount of detail in the magnetic map depends on 
unpolarised line width, projected rotational velocity \(v \sin i\),
signal-to-noise ratio of the observations, and (implementation dependent) choices of fitting parameters such as the \(\chi^2\) target values used in~\citetalias{2016MNRAS.457..580F,2018MNRAS.474.4956F}, see the discussion in~\citet{1997A&A...326.1135D,2010MNRAS.407.2269M}. 

Modern ZDI gives a parametric representation of the vector surface magnetic field \(\vec B(\theta,\varphi)\) in terms of spherical harmonics coefficients~\citep{1999MNRAS.305L..35J,2006MNRAS.370..629D}. Here \(\theta\) and \(\varphi\) denote the polar and azimuth position on the stellar surface in a co-rotating spherical coordinate system (similar to colatitude and longitude on Earth).
For the magnetograms in this study, the surface radial magnetic field \(B_r(\theta, \varphi)\), is parametrised by a set of complex-valued harmonic coefficients \(\alpha_{\ell m}\) and calculated via a spherical harmonics expansion as 
\begin{equation}\label{eq:zdi_br}
    B_r(\theta, \varphi) =
    \operatorname{Re}
    \sum_{\ell=1}^{\ell_\text{max}} \,
    \sum_{m=0}^\ell 
    \alpha_{\ell m} 
    \sqrt{\frac{2\ell+1}{4\pi} \frac{(\ell-m)!}{(\ell+m)!}} 
    P_{\ell m}(\mu) e^{i m\varphi}
\end{equation}
where \(\operatorname{Re}\) denotes taking the real part, \(P_{\ell m}(\mu) \) is the associated Legendre polynomial of order \(m\) and degree \(\ell\), and  \(\mu=\cos\theta\).  Since we discard the imaginary component of the right hand side in equation \eqref{eq:zdi_br} the non-negative orders \(m\geq 0\) provide sufficient degrees of freedom to represent any magnetic field configuration. 
The maximum degree \(\ell_\text{max}\) is set to \num{15}, giving a minimum angular resolution of \(\SI{180}{\degree}/\ell_\text{max} = \SI{12}{\degree}\) in the magnetograms.
 
When comparing magnetograms with similar \(\ell_\text{max}\) values, it is important to note that \(\ell_\text{max}\) is only an upper bound on the representable magnetogram complexity. For slow rotators there may not be sufficient signal in the ZDI profile to reconstruct features past the first few degrees, as the signal scales with \(v \sin i\)~\citep{2010MNRAS.407.2269M} and thus decreases with increasing \(P_\Rot\). To quantify the physical complexity of the magnetograms we calculate an `effective degree' \(\ell_{.90}\) and \(\ell_{.99}\)~\citepalias[see][]{paper2} for which respectively \SI{90}{\percent} and \SI{99}{\percent} of the wind model magnetic energy is contained in degrees smaller than or equal to \(\ell\). The effective degree of the magnetograms in Fig.~\ref{fig:magnetograms-zdi} is discussed in Section~\ref{sec:coronal-magnetic-field}.

While there is general agreement that ZDI is able to reconstruct the large-scale polarity structure of the field~\citep[e.g.\ ][]{2000MNRAS.318..961H}, uncertainty remains surrounding the field strength~\citep{2019MNRAS.483.5246L}. When comparing with the complementary Zeeman broadening technique, which is sensitive to both the large- and small-scale field, \citet{2015ApJ...813L..31Y} found only about \SI{20}{\percent} of field to be reconstructed with ZDI.\@ In this work we control for this possible underreporting of the large-scale magnetic field strength by creating two series of wind models, one of which has its surface magnetic field increased by a factor of 5; we return to this in Section~\ref{sec:numerical_model} and in the analysis in Section~\ref{sec:magnetic-scaling}.

\section{Simulations}\label{sec:Simulations}
We\label{sec:model-equations} obtain the wind models of this work by numerically solving a two-temperature extended ideal MHD model with physical heating and cooling terms, and letting the models evolve towards a steady-state solution. For this we use the  Alfvén wave Solar model
\cite[\awsom{}, ][]{2013ApJ...764...23S, 2014ApJ...782...81V} driven by the surface magnetic fields in Fig.~\ref{fig:magnetograms-zdi}. 
In the \awsom{} model the mechanism of coronal heating is Alfvén wave dissipation~\citep[see e.g.\,][]{1968ApJ...154..751B}. 
The \awsom{} model was created to model the Solar wind, in which context it has been the subject of extensive 
validation~\citep{2015MNRAS.454.3697M,2019ApJ...872L..18V,2019ApJ...887...83S}. The model is also used in stellar winds modelling~\citep[e.g.\ ][and \citetalias{paper1,paper2}]{2016A&A...588A..28A,2016A&A...594A..95A,2017ApJ...835..220C,2017ApJ...843L..33G,2021MNRAS.504.1511K}. 
The \awsom{} model is part of the block-adaptive tree solarwind Roe upwind scheme~\citep[\batsrus{}, ][]{1999JCoPh.154..284P,2012JCoPh.231..870T} code, which in turn is part of the space weather modelling framework 
\cite[\swmf{}, ][]{2005JGRA..11012226T,2012JCoPh.231..870T,2021JSWSC..11...42G}. We refer the reader to \citetalias{paper1,paper2} for a detailed description of the model equations and modelling parameters used in this work; here we provide a brief summary of the coronal heating mechanism in the \awsom{} model and the magnetic field scaling applied to compensate for the issues considered in Section~\ref{sec:mag-map-zdi}.

As the \awsom{} module extends outwards from the stellar chromosphere, an Alfvén wave energy flux is prescribed at the inner boundary. The Alfvén wave energy flux is modelled as a boundary Poynting flux-to-field ratio \(\Pi_\Alfven / B\), such that the local amount of Alfvén wave energy crossing the inner model boundary is proportional to the local value of the magnetic field \(|\vec B(\theta, \varphi)|\).  We note that the parameters used in this paper and \citetalias{paper1,paper2} have been chosen to match Solar values and reproduce Solar conditions based on Solar magnetograms. As such, they are likely to be more accurate the closer the age of the modelled star is to the age of the Sun, and the values may be more questionable for the \(\SI{\leq 0.13}{\giga\year}\) young stars of this study compared to the older stars in~\citetalias{paper1,paper2}. We also consider this issue in Section~\ref{sec:Conclusions}.

In\label{sec:numerical_model}  our wind models the numerical domain extends from \numrange{1}{400} stellar radii, split into an inner domain using a spherical grid and an outer domain (beyond \num{45} stellar radii) using a Cartesian grid. We first find a steady state on the inner grid, then we use this solution to initialise the outer grid. For the outer grid we apply a local artificial wind flux scheme~\citep{2002JCoPh.181..354S} with an mc3 slope limiter~\citep{bfe87ab1e06744e7b6e2163728ba7ac4}
as we did in~\citetalias{paper1,paper2}. 

For the inner, spherical domain we now apply an HLLE scheme as modified by~\cite{https://doi.org/10.1002/fld.312} and a minmod type slope limiter~\citep{1986AnRFM..18..337R}. 
To incorporate electron heat conduction~\citepalias[see][]{paper1} we apply a semi-implicit time stepping method, which effectively treats the thermal conduction operator implicitly~\citep{1998A&A...332.1159T,1999IJNMF..30..335K}. Furthermore, we have extended the inner domain with an outer buffer region 
in order to 
prevent unphysical inflow situations 
from destabilising the solution before a steady state is reached. 
We apply `outflow' boundary conditions at the outer edge of the buffer region with an outflow pressure of \SI{e-12}{\pascal}. 
This minimises unphysical inwards-propagating disturbances caused by sub-Alfvénic (see Section~\ref{sec:alfven-surface-and-current-sheet}) flows across the outer boundary (these can form in the early stages of a model run but are not present in the final solution). We emphasise that the final, steady-state solutions do not exhibit destabilising sub-Alfvénic flow at outer boundary in any of our models.

In this work all the wind models have been created with the open source version of the SWMF\footnote{The open source version of the space weather modelling framework can be found on-line at 
    \url{https://github.com/MSTEM-QUDA/SWMF.git}. In this work version \href{https://github.com/MSTEM-QUDA/SWMF/commit/a5beb110f}{2.40.{\tt a5beb110f}} (2022-10-08) has been used for all processing.
}.
The options discussed here are all available via configuration parameters. The application of the above changes to our wind modelling approach has enabled us to model the winds of the young, rapidly rotating stars in Table~\ref{tab:observed_quantities}. 
We have also re-processed our previous wind models~\citepalias[from][]{paper1,paper2} with this new and stable numerical configuration, so that the same methodology and model configuration is applied across the whole age range from \SIrange{24}{650}{\mega\year}. The re-processing of our previous wind models has led to only minor differences, and thus we do not present the re-processed wind models in Section~\ref{sec:results}; we focus instead on our modelling of the stars of Table~\ref{tab:observed_quantities}. In Section~\ref{sec:Discussion} onwards we do, however, use the aggregate wind parameters computed from the re-processed wind models and the models in this work, so that we have a fully consistent modelling methodology. The aggregate wind parameters of the re-processed wind models are also given in Appendix~\ref{sec:full-table}.

In an attempt to control for ZDI's inherent uncertainty concerning the magnetic field strength (see Section~\ref{sec:mag-map-zdi}), we create two wind models for each star corresponding to different scalings of the surface magnetic field maps of Fig.~\ref{fig:magnetograms-zdi}. 
The \(B_\ZDI\) series uses the unscaled magnetic field of equation~\eqref{eq:zdi_br}, while \(5B_\ZDI\) series applies a scaling factor of 5 to the \(B_\ZDI\) values  (this approach was also used in~\citetalias{paper1} and~\citetalias{paper2}). 
In this work we refer to each wind model by its case name of the form `\C1F' or `\C5F'; these refer to the star HII~296 from Table~\ref{tab:observed_quantities} with the \(B_\ZDI\) and \(5B_\ZDI\) surface magnetic fields respectively. The symbol shape indicate whether the case belongs to  \(B_\ZDI\) or the \(5B_\ZDI\) series, and its colours indicate the star's association (see Table~\ref{tab:observed_quantities}).

\section{Results}\label{sec:results}
In this section we present the wind model results and aggregate quantities that may be calculated from the wind model output. We present plots and key quantities describing the final, steady-state coronal magnetic field, the wind speed and Alfvén surface location, the wind mass loss rate and angular momentum loss rate, and plots of the wind pressure in the equatorial plane extending out to planetary distances.

\subsection{Coronal magnetic field}\label{sec:coronal-magnetic-field}

In this section we describe the final, steady-state coronal magnetic field in our wind model solutions where the hydrodynamic, magnetic and thermal forces are in balance inside our model domain.

Table~\ref{tab:magnetic_averages} gives aggregate magnetic quantities calculated from the steady-state wind models. These quantities are calculated at the inner boundary of the wind model. The quantity \(|B_r|\) is the average value of the unsigned radial field over the stellar surface, whereas \(\max |B_r|\) is similarly the maximum value of the radial field over the stellar surface. The quantities \(|B_r|\) and \(\max |B_r|\) can also be calculated from equation \eqref{eq:zdi_br} and the two methods are in agreement.

\begin{table*}
    \centering
    \caption{
        Aggregate magnetic quantities for the unscaled \(B_\ZDI\) and scaled \(5B_\ZDI\) magnetic fields considered in this work. The `Case' column gives the star's case name, which is prepended by `\(5\times\)` for models in the \(5B_\ZDI\) series. An identifying symbol as in Table~\ref{tab:observed_quantities} for models in the \(B_\ZDI\) series and in the shape of a star for models in the \(5B_\ZDI\) series is also provided for each model, and used throughout this paper. In the following columns 
        \(|B_r|\) and \(\max |B_r|\) are the mean and maximal unsigned radial magnetic surface field strength, calculated from the magnetograms in Fig.~\ref{fig:magnetograms-zdi}. \(|\vec B|\) is the average surface magnetic field strength of the final, steady-state wind models in Fig.~\ref{fig:br-fieldlines}.
         \(\Phi = 4\pi R_\Star^2 |B_r|\) is the unsigned magnetic flux at the stellar surface. The `dip.', `quad.' `oct.' and `16+' columns give the proportion of magnetic energy associated with respectively the dipolar (\(\ell=1\)), quadrupolar (\(\ell=2\)), octupolar (\(\ell=3\))
    and higher order (\(\ell \geq 4\)) degrees of the spherical harmonics coefficients \(\alpha_{\ell m}\) in equation \eqref{eq:zdi_br}. The \(\ell_{0.90}\) and \(\ell_{0.99}\) columns give the magnetogram degree for which \SI{90}{\percent} and \SI{99}{\percent} of the magnetic energy is contained in spherical harmonics of degree less than or equal to \(\ell\). 
    }\label{tab:magnetic_averages}
    \sisetup{
        table-format=2.1,
        table-figures-decimal=1,
        table-figures-integer=2,
        table-number-alignment=center,
        round-mode=places,
        round-precision=1
    }
    \input{tables/aggregate-magnetic-quantities.tex}
\end{table*}
\begin{figure*}
    \centering
    \includegraphics{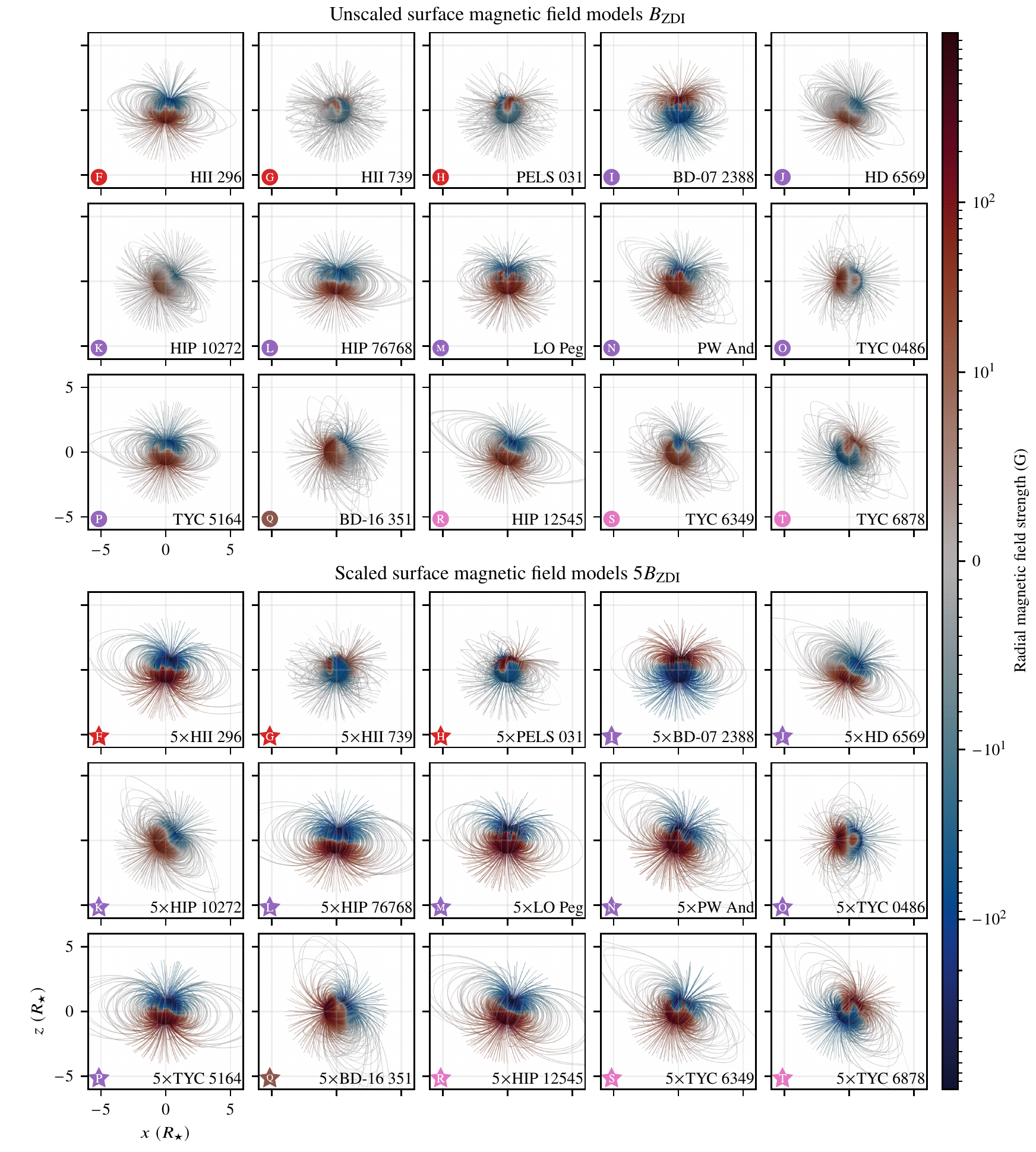}
    \caption{
        Coronal magnetic field of the steady-state wind solution found using \awsom{}. 
        Open magnetic field lines are truncated at \(4R_\Star\). The surface magnetic field of Fig.~\ref{fig:magnetograms-zdi} dominates in the inner regions while the wind's pull on the magnetic field lines becomes more important in the  outer regions. This can be observed where closed field lines are pulled into egg-shaped and/or pointy `helmet streamer' shapes.
        The field lines and the stellar surface are coloured according to the colour scale; note that the scale is linear between \SIrange[]{-10}{10}{\gauss} and logarithmic outside of this region. Many of the coronal fields are largely dipolar, characterised by two regions of open field lines with opposing polarity and a region of closed field lines near the magnetic equator. We display each star so that the dipolar structure is evident by selecting the stellar phase of rotation. With the exception of \C1G, \C1H and their scaled counterparts, the stars exhibit dipole-like coronal magnetic fields with a range of orientations relative to the stellar spin axis \(\uvec \Omega \parallel \uvec z\).  
    }\label{fig:br-fieldlines}
\end{figure*}

Continuing with the quantities of 
Table~\ref{tab:magnetic_averages},
the quantity \(|\vec B|\) is the final, relaxed average (vector) magnetic field strength over the stellar surface in the wind models. The vector field 
\(\vec B = B_r \uvec r +\vec B_\perp \) 
will relax towards a steady-state, and the non-radial magnetic field \(\vec B_\perp\) at the stellar surface is free to settle since only the radial field \(B_r\) is fixed by a boundary condition. As the \batsrus{} \awsom{} code is driven with the radial magnetic field \(B_r\) only, \(\vec B_\perp\) is determined by the physics above the chromosphere. 
This is in contrast with the ZDI magnetic field, where the non-radial magnetic field is also affected by photospheric currents. As the photosphere lies below the chromosphere it is not part of the \awsom{} model; we instead observe that the relaxed values of \(\vec B\) are close to the potential component of the ZDI magnetic field. The potential component of the ZDI magnetic field is thought to determine the magnetic geometry of the corona, while the non-potential field may be stirred by rapid stellar rotation. The ZDI field component originating from photospheric currents is not thought to affect the wind to a significant degree~\citep{2013MNRAS.431..528J}. 

The quantity \(\Phi\) in Table~\ref{tab:magnetic_averages} is the unsigned surface magnetic flux. The unsigned magnetic flux through a closed surface is 
\begin{equation}\label{eq:unsigned-flux}
    \Phi(S) = \oint\nolimits_{S} |\vec B \cdot \uvec n| \, \mathrm{d}S,
\end{equation}
where \(B\) is the magnetic field and \(\uvec n\) is the normal of the surface \(S\).
The surface magnetic flux \(\Phi(S_\Star)\) at the inner boundary \(S_\Star\) of our model is given in 
Table~\ref{tab:magnetic_averages}. This quantity can also be found from the equivalent \(\Phi(S_\Star)=4\pi R_\Star^2 |B_r|\) with excellent agreement between the two methods.

The remaining columns of Table~\ref{tab:magnetic_averages} give several measures of the surface magnetic field complexity in each wind model. The `Dip.', `Quad.', `Oct.' and `16+' columns give the fraction of magnetic energy contained in magnetogram terms of degree \(\ell=1\), \(\ell=2\), \(\ell=3\), and \(\ell\geq 4\), respectively, i.e.\ the amount of energy in the dipolar, quadrupolar, octupolar, and hexadecapolar and higher degrees. Here we also see good agreement with the values of~\citetalias{2016MNRAS.457..580F,2018MNRAS.474.4956F}. In Section~\ref{sec:mag-map-zdi} we mentioned that the magnetogram `effective degree' may lie significantly below the maximum \(\ell_\text{max}\) of the spherical harmonics expansion in equation~\eqref{eq:zdi_br} when most of the magnetic energy is contained in spherical harmonics terms of low degree. In the last two columns of Table~\ref{tab:magnetic_averages} we provide two measures of the effective degree, \(\ell_{.90}\) and \(\ell_{.99}\), which are the magnetogram degrees \(\ell\) for which \SI{90}{\percent} and \SI{99}{\percent} of the surface magnetic field energy is contained in degrees less than or equal to \(\ell\). It may be seen that the rapid rotators
\C1I and \C1M
are the only stars with \(\ell_{.99}=\ell_\text{max}\) in our sample. The second highest effective degree is found in 
\C1N 
with \(\ell_\text{.99}=12\), while the rest of the stars in our sample range from \(\ell_{0.99}=4\) 
(for \C1J) 
to \(\ell_{.99}=10\). A similar variation is seen for the calculated \(\ell_{.90}\) values. These variations in level of reconstructed detail can be recognised in Fig.~\ref{fig:magnetograms-zdi} when comparing the radial magnetic fields of e.g.\  \C1I and \C1J. 

Fig.~\ref{fig:br-fieldlines} shows the steady-state coronal magnetic field structure for the wind models created in this work. Each panel of the plot shows the coronal magnetic field structure by tracing a large number of magnetic field lines from an evenly sampled set of points on the stellar surface. Due to the presence of hydrodynamic forces in the wind solution,
magnetic field lines can be dragged along with the wind velocity field when the hydrodynamic pressure exceeds the magnetic pressure. 
In our simulations these magnetic field lines terminate at the outer boundary of the wind model; they are termed `open' since they have only one footpoint on the stellar surface. 
For display purposes the open field lines are truncated at four stellar radii, while closed magnetic field lines, which connect two different points on the stellar surface, are not truncated. The stellar surface and the magnetic field lines are coloured according to the local value of the radial magnetic field strength \(B_r\); we use a colour scale which is linear from \SIrange{-10}{10}{\gauss} and logarithmic outside of this range. 

Considering a closed spherical surface \(S\) of radius \(R\) centred on the star, the `open magnetic flux' is the part of the magnetic flux associated with regions of \(S\) crossed by open magnetic field lines. The closed magnetic flux is, vice versa, the amount of magnetic flux in regions of \(S\) where the magnetic field lines close back on \(S\). 
Splitting the magnetic flux \(\Phi(S)\) through \(S\) into an open and a closed component \(\Phi(S) = \Phi_\Open(S) + \Phi_\text{closed}(S)\) it can be seen that \(\Phi_\text{closed}(S)\) tends towards zero for \(R \gg R_\Star\). We use this to calculate open magnetic flux \(\Phi_\Open\) by evaluating \(\Phi(S)\) for \(R \gg R_\Star\) where \(\Phi_\Open \gg \Phi_\text{closed}\). The values we find are consistent with the above argument; we give the open flux values as \(\Phi_\Open\) in Table~\ref{tab:main-table}. 
\begin{table*}
    \centering
    \caption{
        Aggregate parameters calculated from the wind models of this work. All the quantities here are calculated for the final, steady-state winds.
        \(\Phi\) is the magnetic flux at the stellar surface,
        \(\Phi_\Open\) is the surface flux for which the field lines are open,
        \(S_\Open\) is the area of the stellar surface crossed by open magnetic field lines,
        \(i_\CS\) is the inclination of the inner current sheet relative to the stellar axis of rotation \(\uvec \Omega\),
        \(\Phi_\text{axi}\) is the axisymmetric component of the open flux \(\Phi_\Open\),
        \(R_\Alfven\) is the average Alfvén radius,
        \(|\vec r_\Alfven \times \uvec \Omega|\) is the cylindrical Alfvén radius,
        \(\dot M\) is the wind mass loss rate,
        \(\dot J\) is the wind angular momentum loss rate,
        \(P_\Wind^\Earth\) is the average ambient wind pressure for an Earth-like planet,
        \(R_\Mag\) is the average magnetospheric stand-off distance for an Earth-like planet.
        In the case that \(\vec r_\Alfven\) extends past the model domain this precludes the calculation of \(R_\Alfven\) and \(|\vec r_\Alfven \times \uvec \Omega|\); this is indicated by a dagger (\(\dagger\)) symbol in the `Case' column.
    }\label{tab:main-table}
    \sisetup{
        table-figures-decimal=1,
        table-figures-integer=1,
        table-figures-uncertainty=0,
        table-figures-exponent = 3,
        table-number-alignment=center,
        round-mode=places,
        round-precision=1
    }
    \input{tables/main-table-body.tex}

\end{table*}

Table~\ref{tab:main-table} also gives the area of the stellar surface that is crossed by open magnetic field lines in the \(S_\Open\) column. This quantity is found by tracing the evenly spaced magnetic field lines shown in Fig.~\ref{fig:br-fieldlines} until the field line either reaches the edge of the computational domain (open field line), or loops back to the stellar surface (closed field line). \(S_\Open\) is then the number of open field lines divided by the total number of field lines. From Table~\ref{tab:magnetic_averages} we can see that \(\Phi_\Open\) and \(S_\Open\) are well correlated 
with \(S_\Open/S \sim \num{0.6}\, \Phi_\Open/\Phi\).
The complex coronal magnetic fields of 
\C1G
and
\C1H (in Fig.~\ref{fig:br-fieldlines})
do not seem to produce lower values of \(\Phi_\Open\) and \(S_\Open\) than the rest of the sample. 

In most of our model cases %
we observe dipole-dominated coronal magnetic fields, except 
\C1G,  
\C1H, and their scaled counterparts, for which the coronal magnetic fields appear more complex. 
Indeed the amount of dipolar magnetic field energy of these two stars (in Table~\ref{tab:magnetic_averages}) is the lowest of our sample at \SI{25}{\percent} and \SI{10}{\percent} respectively.
The stars \C1I and \C1M are notable as they have a very high effective magnetogram degree measure \(\ell_{0.99} = 15\); this does not, however, stop them from forming a dipole-like coronal magnetic field as both stars have \SI{\sim 40}{\percent} of their magnetic energy in the dipolar terms of the magnetogram. \C1O also exhibits a dipole-like coronal field with \SI{37.9}{\percent} of its magnetic energy in dipolar magnetogram terms. This suggests that the lower threshold of dipolar magnetic energy for forming a dipolar coronal field lies at around \SI{\sim30}{\percent} of the total magnetic energy in our wind models.

We observe a range of `magnetic inclinations', measured as the  angle between the stellar axis of rotation \(\uvec \Omega \parallel \uvec z\) and the `north and south pole' of the magnetic dipole in Fig.~\ref{fig:br-fieldlines}; in order to quantify this we calculate magnetic inclination in the corona by fitting a plane to the surface where \(B_r = 0\). Letting \(\uvec n_\CS\) be the normal vector of this plane, we compute the magnetic inclination \(i_\CS = \cos^{-1}\left(\uvec \Omega \cdot \uvec n_\CS\right)\). The values of \(i_\CS\) are given in Table~\ref{tab:main-table}; we see that 
\C1F,
\C1I,
\C1L, 
and
\C1P have \(i_\CS \lesssim \SI{15}{\degree}\); these stars also have low magnetic inclinations in Fig.~\ref{fig:br-fieldlines}, with the exception of \C1F, which does not exhibit a dipole-dominated coronal magnetic field. \C1O and 
\C1Q, which have values of \(i_\CS \gtrsim \SI{70}{\degree}\) are seen to be highly magnetically inclined in Fig.~\ref{fig:br-fieldlines}. High magnetic inclinations and rapid stellar rotation create more complex wind geometries (see Section \ref{sec:alfven-surface-and-current-sheet}).

The final coronal field quantity we calculate is the axisymmetric open magnetic flux, which has been linked to the amount of cosmic rays propagating in the inner Solar system and impacting the Earth
~\citep{2006ApJ...644..638W} following~\citet{2014MNRAS.438.1162V}:
\begin{equation}
    \label{eq:axisymmetric-flux}
    \Phi_\text{axi} = \oint\nolimits_S |\vec B_\text{axi} \cdot \uvec n| \,\mathrm{d}S, 
    \text{ where }
    \vec B_\text{axi} = \frac{1}{2\pi}\oint \vec B \,\mathrm{d}\varphi. 
\end{equation}
Here \(\vec B_\text{axi}\) is to be understood as the azimuthally averaged value in spherical coordinates \(\vec B = B_r\uvec r + B_\theta\uvec\theta + B_\varphi \uvec \varphi\). We obtain the tabulated values of \(\Phi_\Open\) by integrating over a large spherical surface \(S\) with radius \(R\gg R_\star\), similarly to the calculation of \(\Phi_\Open\). We note that there is a close correlation between \(\Phi_\text{axi} / \Phi_\text{open}\) and \(i_\textsc{B}\); this is expected as both parameters measure the lack of magnetic symmetry around the star's rotation axis.

\subsection{Alfvén surface, wind velocity and current sheet}\label{sec:alfven-surface-and-current-sheet}
Alfvén waves propagate with a speed \(u_\Alfven = B / \sqrt{\mu_0 \rho}\) where \(\rho\) is the wind mass density and \(\mu_0\) is the magnetic constant. Alfvén waves can thus not propagate upstream against the flow velocity \(\vec u\) when the flow speed \(|\vec u| > u_\Alfven\).  As the wind travels away from the stellar surface, the local magnetic field strength and wind density drops, while the wind speed increases. This leads the local wind speed \(u\) to eventually exceed \(u_\Alfven\); 
the set of points where this first occurs forms a closed surface surrounding the star. 
This closed surface \(S_\Alfven\) is called the Alfvén surface; it separates the inner sub-Alfvénic region from the outer super-Alfvénic region of the wind model. Since the wind flow is nearly radial, the sub-Alfvénic region is disconnected from the super-Alfvénic region in the sense that no Alfvén wave disturbance will cross the Alfvén surface inwards (in the \(-\uvec r\) direction). 

The inability of Alfvén waves to propagate inwards suggests that  co-rotating wind cannot effectively drain the star's supply of angular momentum once outside \(S_\Alfven\). This is the conceptual `lever-arm model' of angular momentum loss proposed by~\citet{1962AnAp...25...18S}. While the conceptual lever-arm model is not accurate, it remains useful as the angular momentum loss is strongly affected by the distance to the Alfvén surface; the average Alfvén radius \(R_\Alfven\) is an important parameter in models of stellar spin-down~\citep{1967ApJ...148..217W,1968MNRAS.138..359M,1984LNP...193...49M,1988ApJ...333..236K} which can be written in the form \(\dot J \propto \dot M P^{-1}_\Rot R_\Alfven^n\) where \(n\lesssim 2\) depends on the large scale magnetic structure of the corona, decreasing with increasing field complexity.

The circumstellar current sheet~\citep{1972NASSP.308...44S}, here characterised by \(B_r=0\) is a thin sheet-like region of magnetic reconnection and high currents as magnetic field lines of opposite polarity are forced close together. The condition \(B_r=0\) was also used (in Section~\ref{sec:coronal-magnetic-field}) to calculate the star's magnetic inclination \(i_\CS\). 

\begin{figure*}
    \centering
    \includegraphics{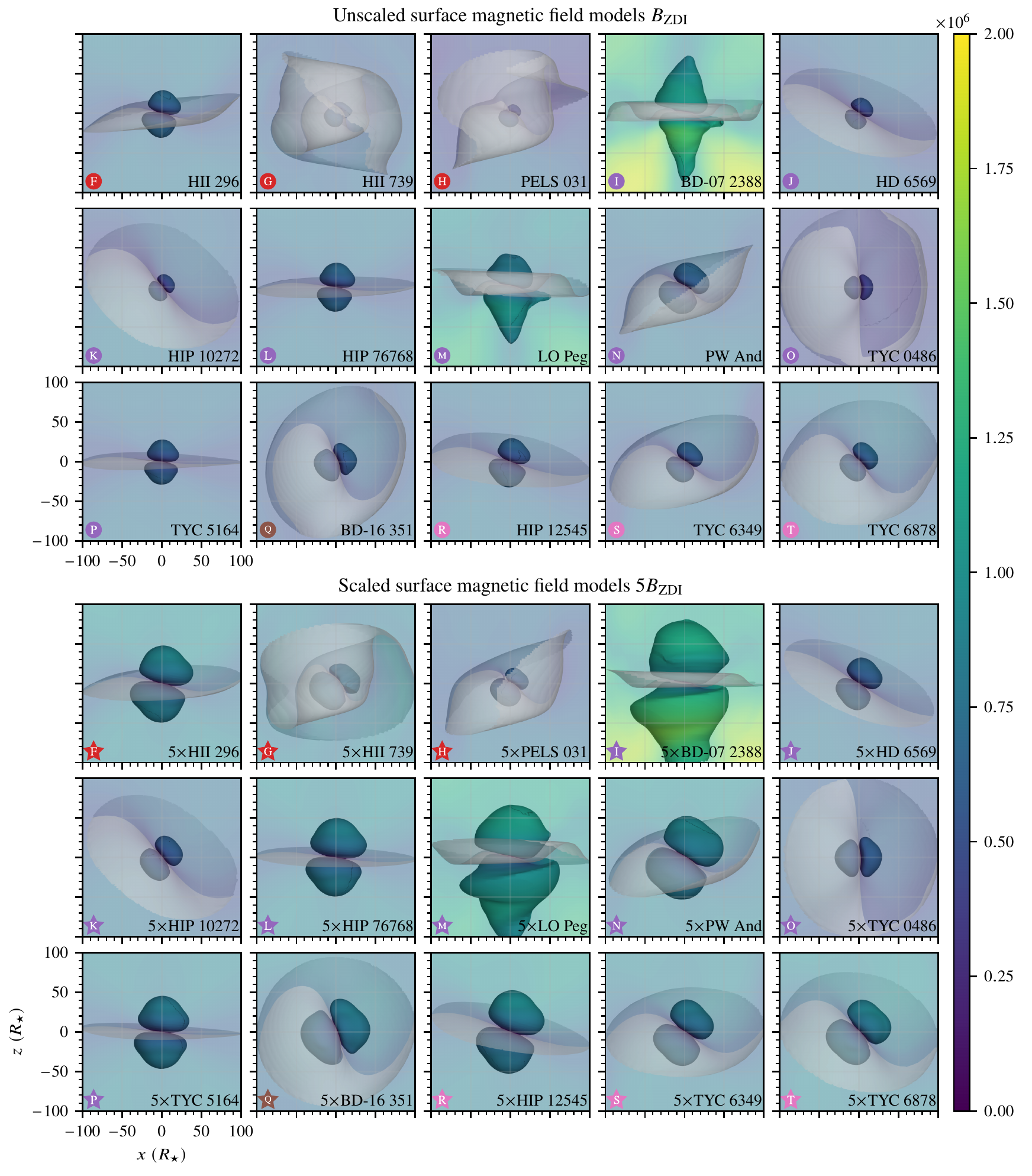}
    \caption{
        Alfvén surface and inner current sheet shown in the same orientation as Fig.~\ref{fig:br-fieldlines}. The \(xz\) plane and the Alfvén surface are coloured by the radial wind velocity. We observe two-lobed Alfvén surfaces independently of the complexity of the inner coronal magnetic field in Fig.~\ref{fig:br-fieldlines}. The inclination of the magnetic dipole to the axis of rotation varies consistently with the dipole patterns seen in the coronal magnetic field. Noticeably, magnetically inclined rapid rotators pull the current sheet around them. The most rapid rotators of the sample, \C1I, \C1M, and their scaled counterparts, exhibit large and sometimes pointy Alfvén surfaces. The pointy shapes arise due to the presence of a significant extended azimuthal magnetic field component in these stars' polar regions; the extended azimuthal field is a consequence of the stars' rapid rotation. 
    }\label{fig:ur-alfven}
\end{figure*}

Fig.~\ref{fig:ur-alfven}  shows the Alfvén surface as an opaque surface and the inner current sheet as a translucent grey surface. The Alfvén surface and the  \(xz\) plane is coloured by the local radial wind speed. We observe two-lobed Alfvén surfaces for every star, including the ones without clear dipolar coronal fields. This shows that in our models only largest scale magnetic features become entrained in the wind. The wind radial velocity and the size of the Alfvén surface lobes appear correlated with the surface magnetic field strength \(|\vec B|\); this is one of many trends we investigate in Section~\ref{sec:Discussion}. 
When the stellar rotation rate \(\Omega\) and the magnetic inclination \(i_\CS\) are high the star's rotation pulls the current sheet around it in a spiraling structure; this is most evident in \C1G and \C1H. 
When the magnetic field exhibits less symmetry around the \(\uvec \Omega\) axis the spiral undulations appear more pronounced \cite[see e.g.\ ][]{2010ApJ...720.1262V}.

For the same star, e.g. \C1G, we see from Fig.~\ref{fig:ur-alfven} that the Alfvén surface tends to lie about a third closer to the star in the \(B_\ZDI\) model series than in the \(5B_\ZDI\) model series. 
The average distance to the Alfvén surface is expected to increases as the surface magnetic field \(B_r\) increases in value~\citep[e.g.\ ][]{1988ApJ...333..236K}; 
we see this increase for all the models although the Alfvén surface of \C1I extends about as far as that of \C5I in the \(+z\) direction. We calculate \(R_\Alfven\) by defining the radial distance to the Alfvén surface \(\vec r_\Alfven(\theta, \varphi)\) over each point \(\theta, \varphi\) on the stellar surface. The average Alfvén radius is then given by 
\begin{equation}
    R_\Alfven = \frac{1}{4\pi R_\Star^2}\oint\nolimits_{S_\Star} \left|\vec r_\Alfven(\theta, \varphi) \right| \, \mathrm{d}S .
\end{equation}
From the resulting values, which are tabulated in Table~\ref{tab:main-table} it is clear that \(R_\Alfven\) increases between the \(B_\ZDI\) and the \(5B_\ZDI\) series.

We also calculate the cylindrical Alfvén radius, this is the mean length of the Alfvén surface's torquing arm around the \(\uvec \Omega\) axis. The value is found by crossing \(\vec r_\Alfven\) with the axis of rotation, \(\vec r_\Alfven(\theta, \varphi)\times \uvec \Omega\), which gives the following expression:
\begin{equation}
    \left|\vec r_\Alfven \times \uvec \Omega \right| 
    = \frac{1}{4\pi R_\Star^2}\oint\nolimits_{S_\Star}  \left|\vec r_\Alfven (\theta, \varphi) \times \uvec \Omega \right| \,\mathrm{d} S
\end{equation}
The calculated cylindrical Alfvén radii are also given in Table~\ref{tab:main-table}. Comparing the growth of \(R_\Alfven\) and \(\left|\vec r_\Alfven \times \uvec \Omega \right| \)  between \C1I and \C5I we see that both quantities have grown by a similar amount, even though the northern Alfvén surface lobe of \C5I appears to have grown mostly thicker and only a bit taller.

Note the unusually large and conical Alfvén surfaces of the very rapidly (\(P_\Rot=\SI{0.3}{\day}\)) rotating \C1I, \C1M (\(P_\Rot=\SI{0.4}{\day}\)) and their scaled counterparts \C5I and \C5M. In the \C5I model the conical southern Alfvén lobe extends past the computational domain so that no value of \(R_\Alfven\) and \(|\vec r_\Alfven \times \uvec \Omega|\) may be computed for this case; this is indicated by dashes in Table~\ref{tab:main-table}. The conical Alfvén surface shapes arise from the presence of a significant azimuthal \(\vec B\)-component which increases the local \(u_\Alfven\) value and thus shifts the local radial position of the Alfvén surface further away from the stellar surface. We note here that the southern hemisphere of stars mapped with ZDI is usually incomplete (see e.g.~\citetalias{2016MNRAS.457..580F}) due to the observing geometry and that the tall southern Alfvén surfaces may not exist in reality.

\subsection{Mass loss and angular momentum loss}\label{sec:mass-loss-and-angular-momentum-loss}

One of the strengths of numerical wind models as opposed to early, analytical models which were often highly symmetrical and/or restricted to the equatorial plane, is the ease of which the fully three-dimensional wind mass loss rate may be calculated. As we already have the density \(\rho\) and the wind velocity \(\vec u\) everywhere in the solution domain we find the wind mass loss rate by integrating over a closed surface \(S\) containing the star:
\begin{equation}\label{eq:mass_loss_integral}
    \dot M = \oint\nolimits_S \rho \vec u \cdot \uvec n \,\mathrm{d} S.
\end{equation}
The wind mass loss of Solar-type stars is difficult to constrain by observations. Based on observations of \num{\sim 16} Solar-type stars~\citet{2004LRSP....1....2W,2005ApJ...628L.143W} found increasing mass loss rates when going backwards in time from the Sun's age to about \SI{\sim 0.7}{\giga\year}, before which they found lower rates of mass loss. The existence of this `wind dividing line' has, however, been called into question~\citep{2021LRSP...18....3V,2021ApJ...915...37W}. While \(\dot M\) is too small to significantly affect the mass of Solar-type stars over their lifetime, the parameter scales the angular momentum loss in semi-empirical models where \(\dot J \propto \dot M P_\Rot^m R_\Alfven^n\).

The wind angular momentum loss, of course, is what slows the star's spin with age. To compute this quantity we follow the general approach of~\citet{1999stma.book.....M,2014MNRAS.438.1162V} but in a slight deviation from~\citetalias{paper1,paper2} (where this quantity was calculated at the Alfvén surface) we calculate the angular momentum loss rate over the same spherical surface \(S\) as used in \eqref{eq:mass_loss_integral}, as this is numerically more convenient for very extended Alfvén surfaces such those of as \C5I and \C5M. The two methods give comparable results. At a spherical surface \(S\) 
\begin{equation}
    \label{eq:angmom_loss}
    \dot J 
    = \oint\nolimits_{S}
    \left(\vec B \times \vec r\right)_{3} \left(\frac{\vec B \cdot \uvec r}{\mu_0} \right)
    + 
    \left(
        \Omega \varpi^2 + \left(
            \vec r \times \vec v
        \right)_3
    \right)
    \left(
        \vec v \cdot \uvec r
    \right)
    \rho
    \, {\rm d} S.
\end{equation}
The scalar angular velocity \(\Omega = 2 \pi / P_\Rot\) is found from Table~\ref{tab:observed_quantities}, 
\(\vec \varpi = \vec{r} - (\vec r \cdot \uvec z) \uvec{z}\) is the cylindrical distance from the origin, and 
\(\vec v = \vec u - \vec \Omega  \times \vec r\) is the wind velocity in the co-rotating frame. The notation \((\cdot)_3\) refers to the \(z\) component of the enclosed vector. The values calculated for \(\dot M\) and \(\dot J\) are given in Table~\ref{tab:main-table}. The \(\Omega\) term in equation \eqref{eq:angmom_loss} suggests that \(\dot J/\Omega\) may vary less than \(\dot J\) itself; we return to this topic in Section~\ref{sec:Discussion}.

\subsection{Wind pressure in the equatorial plane}\label{sec:wind-pressure-in-the-equatorial-plane}
\begin{figure*}
    \centering
    \includegraphics{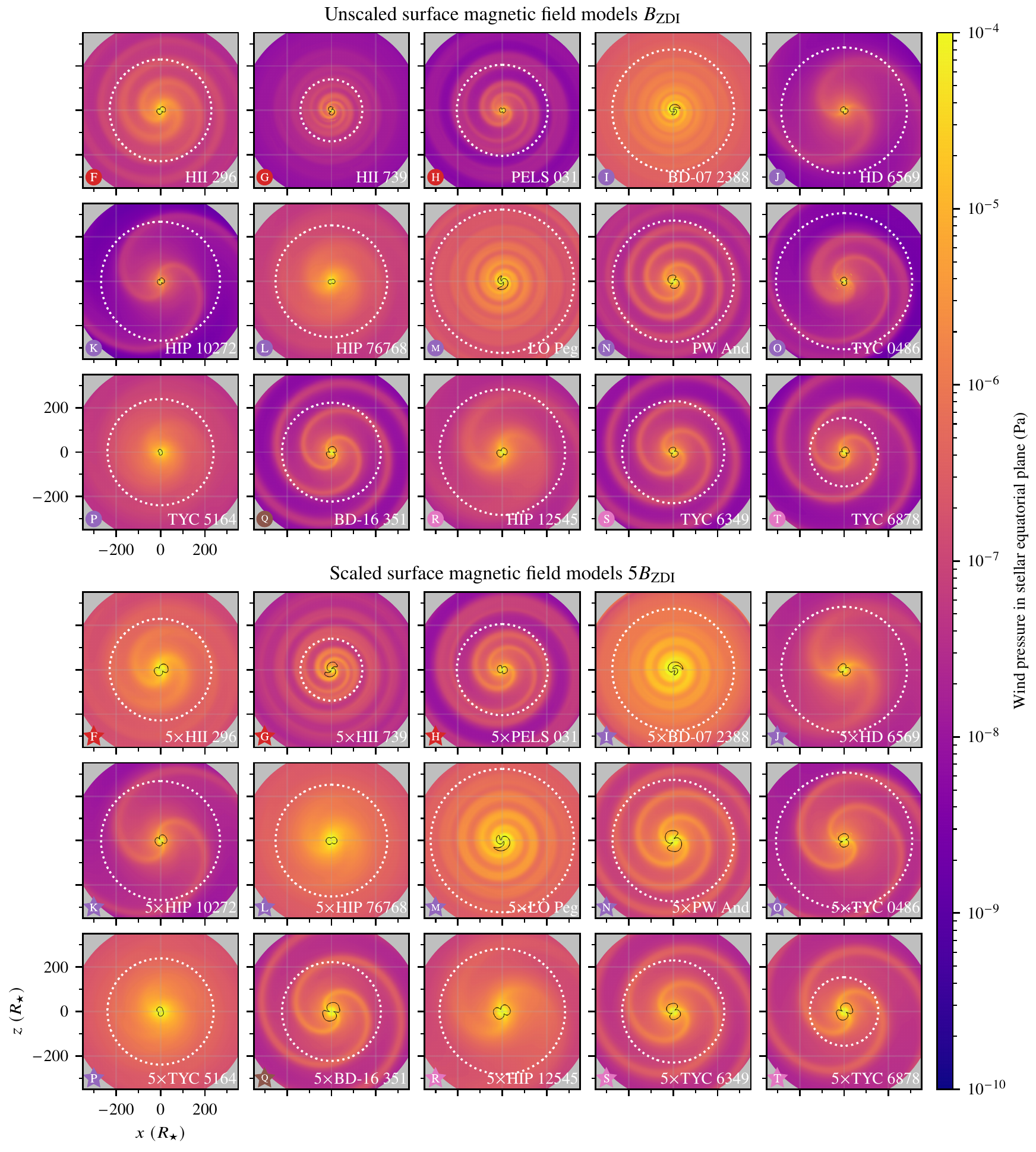}
    \caption{Wind pressure in the stellar equatorial plane, calculated according to equation~\eqref{eq:wind-pressure}. 
    The inclination of the magnetic dipole axis and the geometry of the surface magnetic field gives rise to co-rotating interaction region (CIR) spiral structures with two or more `arms'. The amount of arm winding is governed by the stellar rotation rate and the stellar magnetic field strength. The equatorial cut of the Alfvén surface is indicated by a black line.
    The would-be orbit of an Earth-like planet is indicated by a dotted white line. At and beyond Earth-like distances from the star the wind pressure is dominated by the ram pressure resulting from the wind radial velocity, \(P_\Wind\approx \rho u^2_r\), as the wind velocity is nearly radial and the thermal and magnetic pressure terms drop off rapidly with increasing stellar distance.
     }\label{fig:Pwind-IH-equatorial}
\end{figure*}

The stellar wind pressure exerted on a body travelling with velocity \(\vec w\) through the wind is~\citep{2011MNRAS.414.1573V}
\begin{equation}\label{eq:wind-pressure}
    P_\Wind= P+ |\vec B|^2 / (2 \mu_0) + \rho |\vec u - \vec w|^2,
\end{equation}
here the right hand side terms are the thermal pressure, magnetic pressure and ram pressure respectively. Fig.~\ref{fig:Pwind-IH-equatorial}  shows the wind pressure \(P_\Wind\) exerted on a stationary (\(\vec w=0\)) object in the stellar equatorial plane extending out to the would-be orbit of an Earth-like planet (dotted white line). The intersection of the Alfvén surface (of Fig.~\ref{fig:ur-alfven}) and the stellar equatorial plane is plotted with a black line.
In Fig.~\ref{fig:Pwind-IH-equatorial} we observe as expected that \(P_\Wind\) drops with increasing stellar distance, and that spiral arm-like corotating interaction regions 
\citep[CIRs, see][]{1971JGR....76.3534B,1996ARA&A..34...35G} form when regions of fast stellar wind encounters regions of slow stellar winds. We mainly see two-armed structures corresponding to the dipole-dominated coronal magnetic fields in Fig.~\ref{fig:br-fieldlines}. 
We note that \C1J, \C1L, \C1I, \C1P,  and \C1R, with their low magnetic inclinations, give less pronounced spiral structures than the highly inclined \C1O and \C1Q, with the other models occupying a middle ground. It is clear that the rapid rotators \C1G, \C1I, \C1N and \C1M produce more tightly wound spirals than the other, slower rotators. 

The average wind pressure for an Earth-like planet is tabulated as \(P_\Wind^\Earth\) in Table~\ref{tab:main-table}. This value is calculated by taking the average of equation~\eqref{eq:wind-pressure} over both planetary and stellar phase. At such a large distance from the central star the ram pressure term \(\rho |\vec u- \vec w|^2\) dominates the thermal and magnetic pressure contributions to \(P_\Wind\).
Furthermore the wind is mainly radial \(|\vec u| \approx u_r\), and the wind speed exceeds the planet's orbital speed \(|\vec u| \approx u_r \gg |\vec w|\) so that \(P_\Wind \approx \rho u_r^2\). For reference, monthly \(P^\Earth_\Wind\) averages in the OMNI\footnote{OMNIWeb is available at \url{https://omniweb.gsfc.nasa.gov/}.} \citep[see][]{2005JGRA..110.2104K}  dataset range from \(\mysim 0.1\) to \(\mysim \SI{10}{\nano\pascal}\).

When the super-Alfvénic stellar wind encounters a magnetised body, such as a planet, a standing shock wave will form upstream of the body. 
The magnetospheric stand-off distance \(R_\Mag\) is a measure of the distance from the planet to the shock wave; this distance can be approximated by considering pressure balance  between the wind pressure and the body's dipolar magnetic field strength,
\begin{equation}
    \left.R_\Mag \middle/ R_\text{p}\right. = \left(B_\text{p}^2 \middle/ \left(2 \mu_0 \, P_\Wind\right)\right)^{1/6};
\end{equation}
here \(R_\text{p}\) is the radius of the planet~\citep{1931TeMAE..36...77C,2009ApJ...703.1734V} and \(B_\text{p}\) is the strength of the planet's magnetic field. For our Earth-like planet we use \(B_\text{p}=\SI{0.7}{\gauss}\); this value accounts for magnetospheric currents~\citep{1964JGR....69.1181M}. The calculated values of \(R_\Mag/R_\text{p}\) are given in Table~\ref{tab:main-table}. We see that the magnetospheric stand-off distance ranges from \(3.6 \, R_\text{p}\) to \(8.2 \, R_\text{p}\) which is past the suggested threshold of \({\sim}2\, R_\text{p}\) \citep[of][]{2007AsBio...7..185L} where an Earth-like planet would be protected from atmospheric erosion. As \(R_\text{p}\) increases with stellar age, this suggests that an Earth-like planet would be protected from atmospheric erosion throughout its star's main-sequence lifetime.

\section{Discussion}\label{sec:Discussion}

In this section we examine trends in our results in the magnetic field strength as well as in other parameters. 
In Section~\ref{sec:magnetic-scaling} we study the effect of scaling the magnetic field on the wind mass loss rate and other aggregate quantities. 
In Section~\ref{sec:statistical-trends-and-correlations} we quantify the correlation between magnetic field and wind parameters using ordinary least-squares analysis. In Section~\ref{sec:comparison_with_literture_values} we compare our computed values of wind mass loss and angular momentum loss with literature values.

\subsection{Magnetic field scaling}\label{sec:magnetic-scaling}

\begin{figure*}
    \centering
    \includegraphics{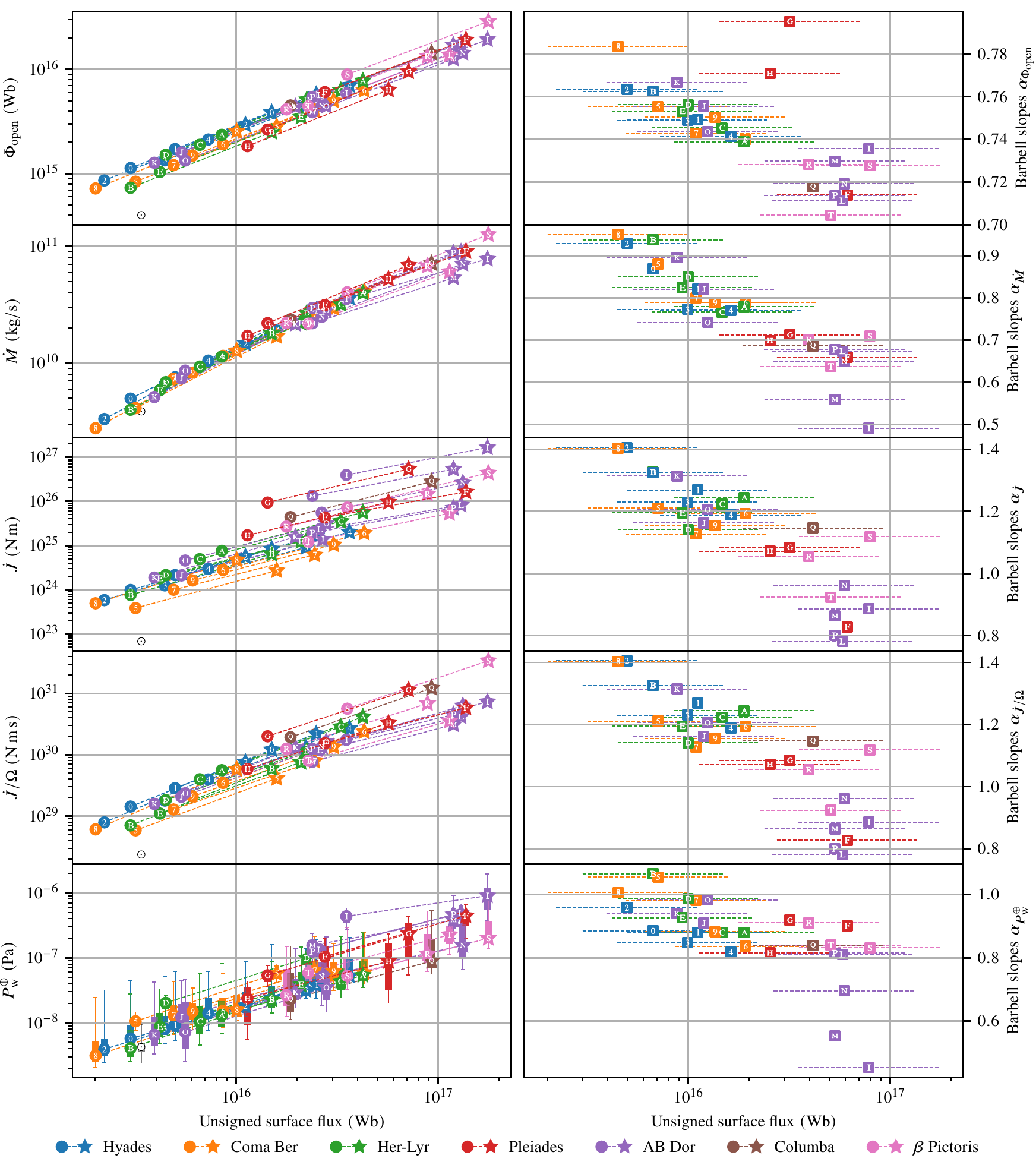}
    \caption{
        Effect of magnetic scaling. On the left hand side, from top to bottom, the open flux, mass loss rate, (rotation scaled) angular momentum loss rate and wind pressure of the models in the \(B_\ZDI\) and \(5B_\ZDI\) series are plotted against the unsigned magnetic flux of eq.~\eqref{eq:unsigned-flux} using the symbols of Table~\ref{tab:magnetic_averages}. For each star, its unscaled and scaled model is connected by a dashed line, yielding a `barbell' shape (also shown in the plot legend).  The Sun symbol represents the Sun at Solar maximum (see~\citetalias{paper1} for details). In the bottom panel the local in-orbit wind pressure variations for an Earth-like planet is represented by boxplots (the lower edge of some boxes are behind the symbol).
        On the right hand side we plot the power-law indices (i.e.\ the \(\alpha_i\) exponents) of each barbell with equation \(\hat y_i(x) = x ^{\alpha_i}\) against the unsigned magnetic flux. For each star we plot a dashed line corresponds to the range between the \(\Phi\) value in the \(B_\ZDI\) and \(5B_\ZDI\) series and a coloured square containing the star's alphanumerical identifier so that each stellar case may be associated with its barbell slope.
    }\label{fig:orthogonal-trend-Unsigned-radial-flux-at-surface} 
\end{figure*} 

As the ZDI reported magnetic field strength is subject to uncertainty, it is of interest to study the effect of pure magnetic field scaling on relevant aggregate quantities of Table~\ref{tab:magnetic_averages} and Table~\ref{tab:main-table}. We carry out such a study by comparing wind models of the same star using the differing magnetic scalings of 1 and 5 times the radial surface magnetic field, i.e.\ by comparing e.g.\ the wind mass loss rate of the 
\C1H and 
\C5H
model cases of the star PELS~031 in Table~\ref{tab:observed_quantities}. Fig.~\ref{fig:orthogonal-trend-Unsigned-radial-flux-at-surface} shows the effect of scaling the magnetic field by comparing the \(B_\ZDI\) series with the \(5B_\ZDI\) series in this way. In addition to the Pleiades, AB~Doradus, Columba, and \(\beta\)~Pictoris stars modelled in this work, we have included models of the Hyades stars from~\citetalias{paper1} and the Coma Berenices and Hercules-Lyra stars from~\citetalias{paper2} (see Appendix~\ref{sec:full-table}). The symbols representing each model case of the same star are linked together by dashed line segments with equations
\begin{equation}\label{eq:scaling-fit}
    \log_{10} \hat y_i(x) = \alpha_i \log_{10} x + \log_{10}\beta_i, \text{ so that } \hat y_i(x) \propto x^{\alpha_i},
\end{equation}
where the index \(i\) ranges over the stars in Table~\ref{tab:observed_quantities}. 
The symbol shapes are the same as is used throughout the paper; a circle for the \(B_\ZDI\) series and a five-pointed star for the \(5B_\ZDI\) series; the colours are the same as in Table~\ref{tab:observed_quantities}. In keeping with~\citetalias{paper1,paper2} we refer to the dashed line segments and symbols of each star as its `barbell'.

While we found tight bounds on the slopes of the fitted lines so that the slopes \(\alpha_i\) had similar values for each star in~\citetalias{paper2}, we observe significantly more spread in the younger stars modelled in this work. The \(B_\ZDI\) series models 
\C1J,
\C1K,  
\C1O,
and their scaled counterparts essentially fall inside the min/max range spanned by the older stars of~\citetalias{paper1,paper2}; 
The remainder of the new stars modelled in this work fall outside~\citetalias{paper2} limits and they do not display a clear linear trend. Instead, we observe a trend where the power-law indices (exponents) \(\alpha_i\) decrease with increasing unsigned surface flux \(\Phi\). To highlight this trend we have plotted the slopes \(\alpha_i\) of each fitted barbell with equation \(\hat y_i(x)\propto x^{\alpha_i}\) in the right hand side of Fig.~\ref{fig:orthogonal-trend-Unsigned-radial-flux-at-surface}. The dashed lines indicate the \(\Phi\) values at the two ends of the barbell, and the star symbols is placed at the geometric mean of the endpoints. 

From the right hand side of Fig.~\ref{fig:orthogonal-trend-Unsigned-radial-flux-at-surface} we see that trend of decreasing \(\alpha_i\) with increasing unsigned surface flux \(\Phi\) is pronounced for the mass loss rate \(\dot M\), the angular momentum loss rate \(\dot J\) and the scaled angular momentum loss rate \(\dot J/\Omega\), where \(\alpha_i\) falls between \num{0.6} and \num{1.4}. The range of variation is smaller for \(\Phi_\Open\) although \C1G and \C1H appear as outliers. Finally, for the wind pressure at an Earth-like planet \(P_\Wind^\Earth\) there is no clear trend of \(\alpha_i\) in \(\Phi\) except that \C1I, \C1M, and \C1N lie below the rest of the model values.

\subsection{Statistical trends and correlations}\label{sec:statistical-trends-and-correlations}
\begin{figure*}
    \centering
    \includegraphics{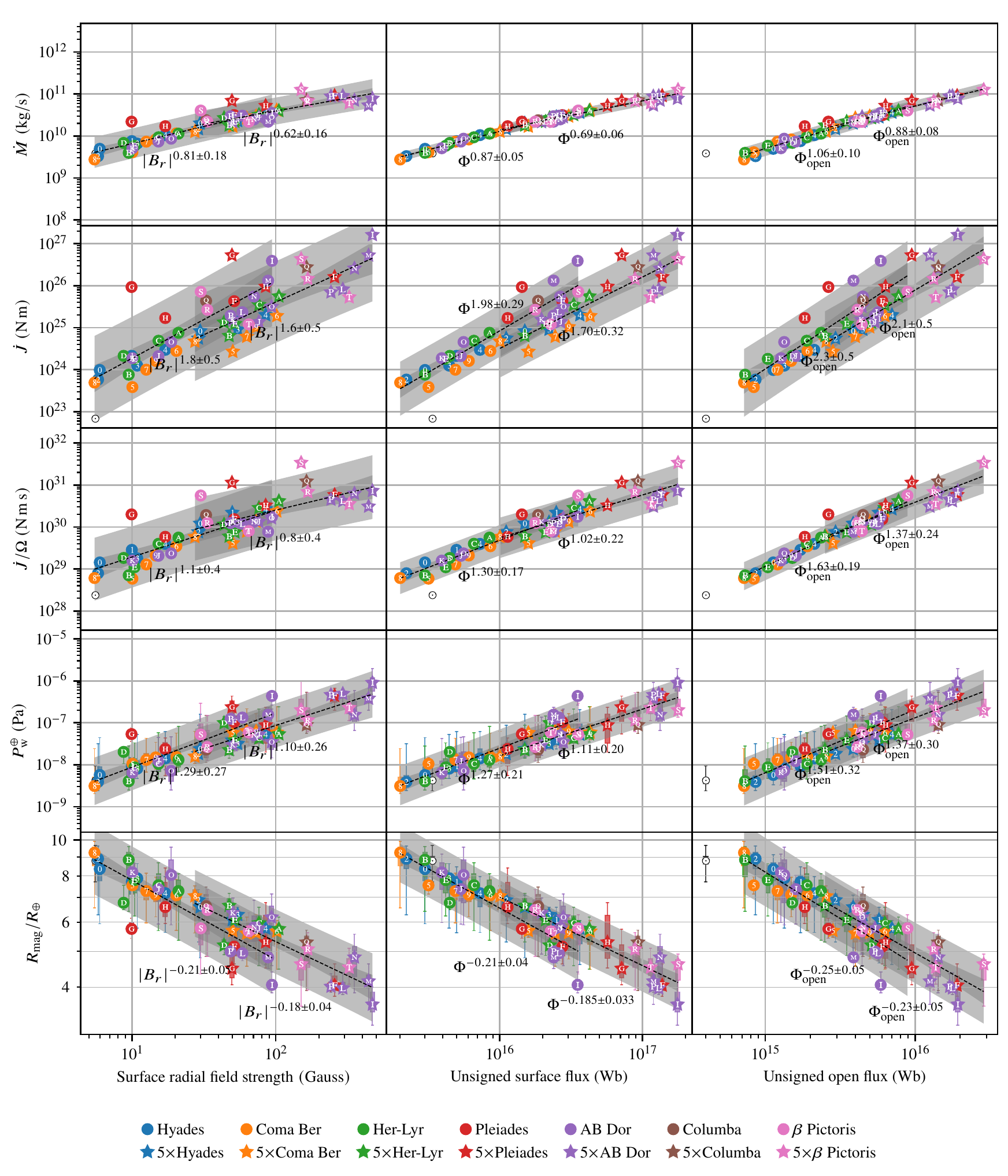}  %
    \caption{
        Variation and trends for the \(B_\ZDI\) and \(5B_\ZDI\) series in several key variables. On the \(x\) axes we plot the average surface radial field strength (left), unsigned surface flux (middle) and unsigned open flux (right). On the \(y\) axes we plot (from top to bottom) mass loss rate, angular momentum loss rate, rotation scaled angular momentum loss rate, wind pressure for an Earth-like planet, and magnetospheric stand-off distance for an Earth-like planet. 
        The coloured symbols correspond to the simulated cases in Table~\ref{tab:magnetic_averages} and Table~\ref{tab:main-table}. For the \(B_\ZDI\) and \(5B_\ZDI\) series we fit a trend line of the form of  equation \eqref{eq:loglog-fit}; the trend lines appear as straight lines as both plot axes are logarithmic. 
        Surrounding the trend line we include  
        confidence bands and prediction bands 
        (dark and light grey regions respectively) at \SI{95}{\percent} confidence level. 
        In the wind pressure and magnetospheric stand-off distance panels the variation along the orbit of an Earth-like planet are indicated by boxplots. Key parameters related to this Figure are given in Table~\ref{tab:correlations-split}. A version of this figure which pools the model cases in \(B_\ZDI\) and \(5B_\ZDI\) series is given in Appendix~\ref{sec:pooled-series}.
    }\label{fig:trend-5x3-broken}
\end{figure*}
\begin{table*}
    \centering
    \caption{
        Summary statistics of the fitted trend lines of the form of equation \eqref{eq:loglog-fit}. Some of these trend lines are shown in Fig.~\ref{fig:trend-5x3-broken}. The top, middle, and bottom part of the table show the correlations of key parameters from Table~\ref{tab:magnetic_averages} and Table~\ref{tab:main-table} with the average surface radial field strength \(B_r\), the unsigned surface magnetic flux \(\Phi\), and the open magnetic flux \(\Phi_\Open\). The parameter of the fitted trend line is given in the quantity column. For the \(B_\ZDI\) and \(5B_\ZDI\) series the fitted power-law index \(a\) and the constant term \(b\) are given along with \SI{95}{\percent} confidence intervals, followed the coefficients of determination \(r^2\), a measure \({y_\text{0.975}}/{y_\text{0.025}}\) of the `height' of the prediction band in Fig.~\ref{fig:trend-5x3-broken} and the probability values \(p\). 
        A version of this table which pools the model cases in \(B_\ZDI\) and \(5B_\ZDI\) series is given in Appendix~\ref{sec:pooled-series}.
        }\label{tab:correlations-split}
    \sisetup{
        table-number-alignment=center,
    }
    \input{tables/correlations-rvalues-pvalues-predint-1-5.tex}
\end{table*}

In this section we carry out a complementary analysis to that of Section~\ref{sec:magnetic-scaling}; instead of considering pairs of models of the same star in the \(B_\ZDI\) and \(5B_\ZDI\) model series, we study the effect of the magnetic field strength, surface magnetic flux, and open magnetic flux on the \(B_\ZDI\) and \(5B_\ZDI\) series separately.

By log-transforming our data the assumptions of ordinary least-squares analysis are satisfied~\citep[see e.g.\ ][]{1998ara..book.....D}. Fig.~\ref{fig:trend-5x3-broken} shows trend lines of the form 
\begin{equation}\label{eq:loglog-fit}
    y(x) = b x^{a} \text{ so that } \log_{10} y(x) = \log_{10} b + a \log_{10} x,
\end{equation}
fitted to the variation of mass loss rate, angular momentum loss rate, rotation-scaled angular momentum loss rate, wind pressure for an Earth-like planet and magnetospheric stand-off distance for an Earth-like planet as a function of the average surface radial magnetic field strength, the unsigned surface magnetic flux, and the open magnetic flux (all these quantities are discussed in Section~\ref{sec:results}).
In each of the panels of Fig.~\ref{fig:trend-5x3-broken}, there are two dashed line segments of the form of equation~\eqref{eq:loglog-fit}, the leftmost one corresponds to the fitted trend line of the \(B_\ZDI\) series, while the rightmost one corresponds to the fitted trend line of the \(5B_\ZDI\) series. Surrounding each fitted trend line are two grey shaded regions; the innermost (dark grey) region is a \SI{95}{\percent} confidence band of the \(a\) and \(b\) parameters of the fitted trend line, while the outermost (light grey) region corresponds to a \SI{95}{\percent} prediction band. The  prediction band variance is the sum of the confidence band variance and the regression mean square error. Given further stars and magnetograms drawn from similar populations as the ones in this study, there is a \SI{95}{\percent} chance of them falling inside these light grey shaded regions. The confidence interval of \(a\) for the \(B_\ZDI\) and \(5B_\ZDI\) series are also printed in the plots, for example we have \(\dot M \propto |B_r|^{0.81\pm0.18}\) for the models of the \(B_\ZDI\) series.

The confidence intervals of \(a\) and \(b\), as well as several other statistical parameters describing the fitted lines of Fig.~\ref{fig:trend-5x3-broken} are given in Table~\ref{tab:correlations-split}.  Using the numerical confidence intervals of \(a\) and \(b\) we can recover the full equation \eqref{eq:loglog-fit} of the fitted lines and confidence bands, e.g.\ 
\(\dot M=\left(10^{9.01\pm0.24}\,\si{\kg\per\second}\right)\left(|B_r|\middle/\SI{1}{\gauss}\right)^{0.81\pm0.18}\) 
for the \(B_\ZDI\) series \(|B_r|\) fitted line. 
Note that when reconstructing the full fitted equations of the form \(y=bx^a\) as in equation~\eqref{eq:loglog-fit}, the choice of physical units for \(x\) and \(y\) will affect the numerical value of \(b\) and its confidence interval. For the numerical values in Table~\ref{tab:correlations-split} all quantities are in SI base units except that the magnetic field is expressed in 
Gauss \(\left(\SI{1}{\gauss} = \SI{e-4}{\tesla}\right)\), 
\(R_\Alfven\) and                             
\(|\vec{r_\Alfven} \times \uvec{\Omega}|\) are expressed in stellar radii (see Table~\ref{tab:observed_quantities}), and \(R_\text{mag}\) is expressed in terms of planetary radii \(R_\text{p}\), as is the case throughout this paper. The magnetic fluxes \(\Phi\) and \(\Phi_\text{open}\) are expressed in Weber (\(\SI{1}{\weber}=\SI{1}{\tesla \square\meter}=\SI{e8}{\gauss\square\centi\meter}=\SI{e8}{Mx}\)).

Continuing with the columns of Table~\ref{tab:correlations-split}, the coefficients of determination (\(r^2\) values) quantify the amount of variation that is explained by the fitted curve, the \({y_{0.975}}/{y_{0.025}}\) values are a physical measure of the amount of variation not explained by the fitted trend line, and the probability values (\(p\) values) give the likelihood that the observed trend is a spurious one that would not exist in a larger sample of stellar models. The confidence intervals, prediction bands, \(r^2\) values, and \(p\) values are calculated in the standard way using the \textsc{statsmodels}~\citep{seabold2010statsmodels} Python package.

The most notable difference between Fig.~\ref{fig:trend-5x3-broken} and the similar analysis of~\citetalias{paper2} is that the \(\Phi\) and \(\Phi_\Open\) ranges are now overlapping between the \(B_\ZDI\) and the \(5B_\ZDI\) model series due to the inclusion of the more powerful and complex magnetic fields of the stars in the AB Doradus moving group, the Pleiades cluster, Columba and \(\beta\) Pictoris, as well as these stars' rapid rotation. In general, we also observe stronger trends with smaller uncertainties (i.e.\ larger power-law indices with reduced error estimates) for the quantities plotted except for \(\dot J/\Omega\) against \(\Phi_\Open\) where the fitted power-law indices are slightly smaller and accompanying error estimates are slightly larger. 
More specifically, in the leftmost column of Fig.~\ref{fig:trend-5x3-broken}, where the quantities are plotted against the mean surface radial field strength \(B_r\), the trend lines from the current full set of models are generally steeper than the trend lines of~\citetalias{paper2} with the exception of \(\dot J/\Omega\) against \(|B_r|\) which is similar to the previous results. This also occurs in the middle column. The fits for \(\dot J/\Omega\) are, however, very similar to those of~\citetalias{paper2}. In the rightmost column, where the model output \(\Phi_\Open\) is on the \(x\) axis, we see that  \(\dot M\) and \(\dot J\) exhibit stronger trends than in~\citetalias{paper2}, similarly to what was seen in the middle column. The fits to \(\dot J/\Omega\) against \(\Phi_\text{open}\) have slightly weaker trends, while the trends in  \(P_\Wind^\Earth\) and \(R_\Mag\) are stronger. 

The model \C1G and its scaled counterpart \C5G are clear outliers in the top three rows of Fig.~\ref{fig:trend-5x3-broken}; in the left column of panels this can be attributable to this star's large radius of \(1.5\,R_\Sun\). \C1G and \C5G remain high in the centre and right columns but their \(x\) position is shifted rightwards because of \(\Phi\)'s dependence on \(R_\Star\).
 The very young star models \C1S and \C5S also exhibit large values of \(\dot J/\Omega\). The other notable outliers, possibly due to their star's rapid \(P_\Rot=\SI{0.3}{\day}\) rotation, are \C1I and \C5I. We do not, however, see a similar effect for the other rapid rotator in the \C1M and \C5M models. 

It is interesting to compare the mass loss rate \(\dot M\) panels in Fig.~\ref{fig:trend-5x3-broken} (top row) where \(\dot M\) is plotted against the average unsigned radial magnetic field strength \(|B_r|\) (left), the unsigned magnetic flux \(\Phi\) (middle) and the open unsigned magnetic flux \(\Phi_\Open\) (right). The four stars \C1I, \C1M, \C1N, and \C1T, and their counterparts in the \(5B_\ZDI\) give the appearance of a weak saturation effect limiting the increase of \(\dot M\) as a function of \(|B_r|\), but the effect is not seen in the panel showing \(\dot M\) as a function of \(\Phi\) (although the four stars still lay below the main trend line). This apparent saturation limiting \(\dot M\)  with increasing \(|\vec B|\) can be seen more clearly in Fig.~\ref{fig:results-context} where we compare the results of our wind modelling with literature values. 
We emphasise to the reader that the apparent saturation limiting \(\dot M\) as a function of increased surface magnetism in our dataset is a spurious effect arising from differences in stellar radius (see Table~\ref{tab:observed_quantities}), and that it is not present when plotting \(\dot M\) against \(\Phi=4 \pi R^2 \, |B_r|\) and thus accounting for where differences in stellar radii.

In Appendix~\ref{sec:pooled-series}, Fig.~\ref{fig:trend-5x3} and Table~\ref{tab:correlations-pooled} show the results of a similar analysis where the model cases in the \(B_\ZDI\) and \(5B_\ZDI\) have been pooled into a single series of sixty stellar wind models. 
\subsection{Comparison with literature values}\label{sec:comparison_with_literture_values}
\begin{figure}
    \centering
    \input{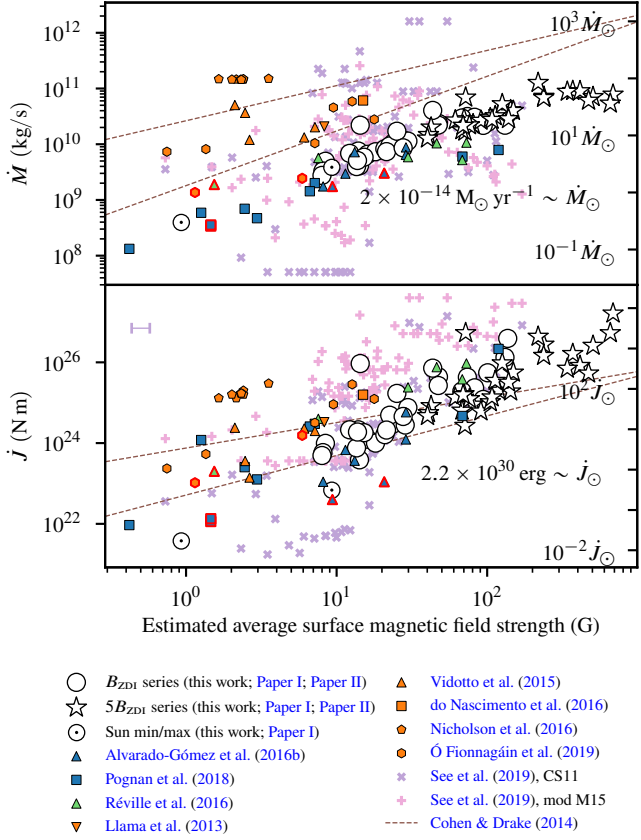}
    \caption{ 
        Comparison of wind mass loss rate \(\dot M\) and wind angular momentum loss rate \(\dot J\) values obtained in this work and from~\citetalias{paper1,paper2} (white symbols) with literature values. The blue symbols denote other stellar wind models created using the \awsom{} model. Orange symbols refer to ideal MHD models where the corona is already hot at the inner model boundary. Symbols with red outlines are Sun models. The plus and cross symbols refer to scaling laws used by~\citet{2019ApJ...886..120S}.
        The brown lines refer to the scaling laws of~\citet{2014ApJ...783...55C} with Sun-like coronal densities and a period of \SI{10}{\day}. We refer the reader to~\citetalias{paper1} for an in-depth explanation of the assumptions behind this plot. 
    }\label{fig:results-context}
\end{figure}

In Fig.~\ref{fig:results-context} we compare the full set of wind mass loss rates and wind angular momentum loss rates described in this work and~\citetalias{paper1,paper2}  with literature values obtained from three-dimensional wind modelling and from scaling laws. 
As the choices and assumptions behind this plot are laid out in~\citetalias{paper1}, we provide this plot here to emphasise that the trends that we observed in~\citetalias{paper1,paper2} still apply. When taken together with the blue symbols representing \awsom{} type three-dimensional wind models, we see an increasing trend in mass loss rate up to about \SI{3e10}{\kilogram\per\second} or 
\num{\sim 24} %
times the fiducial Solar mass loss rate of \SI{2e-14}{M_\Sun \, yr^{-1}}~\citep[e.g.\ ][] {2021LRSP...18....3V} in the \(B_\ZDI\) series of models, and a similar increasing trend up to about \SI{1e11}{\kilogram\per\second} or \num{\sim 80} times the Solar value for the \(5B_\ZDI\) model series. As we noted near the end of Section~\ref{sec:statistical-trends-and-correlations} the apparent threshold of \(\dot M\) values at \(\num{\sim 80} \dot M_\Sun\) is mostly due to stellar radius variation (see Table~\ref{tab:observed_quantities}) in our sample.

Our calculated values of \(\dot M\) are within the spread of the values calculated by~\citet{2019ApJ...886..120S} (see Fig.~\ref{fig:results-context} and see~\citetalias{paper1} for the differences between the CS11 and mod\ M15 methods). We note, however, that in our estimate of mean \(|\vec B|\) from their work (see~\citetalias{paper1}) does not yield mean \(|\vec B|\) values past \SI{\sim 200}{\gauss} such that many of the strongest magnetic fields in the \(5B_\ZDI\) model series have no comparison in their work.
The frequently cited work of~\citet{2002ApJ...574..412W,2005ApJ...628L.143W} also predicts mass loss rates up to \(\num{\sim 100}\dot{M}_\Sun\) 
for \SI{0.7}{\giga\year} old stars, but lower mass loss rates for even younger stars. We do not find strong evidence of a drop in \(\dot M\) values for the youngest stars in our dataset but our results are otherwise well matched to those of their work.

In the bottom panel of Fig.~\ref{fig:results-context} we compare our calculated angular momentum loss rates to those of the literature. 
The right hand side axis shows the angular momentum loss rate in terms of the average Solar angular momentum loss rate, with the caveat that the average Solar wind angular momentum loss rate is not as well constrained as the average Solar wind mass loss rate. We use the observation-based angular momentum loss rate of  \SI{2.2e23}{\newton\meter} from~\citet{2019ApJ...883...67F} as our Solar baseline. 

As can be seen from the red-outlined symbols in Fig~\ref{fig:results-context}, numerical wind models give different \(\dot M\) and \(\dot J\) values depending on the adopted Solar magnetic field strength (i.e.\ phase of the Solar cycle). 
As in \citetalias{paper1,paper2} we continue to find excellent agreement between our models and the \awsom{} based wind models of~\citet{2016A&A...588A..28A} and~\citet{2018ApJ...856...53P}, and good agreement with the models of~\citet{2016ApJ...832..145R}.

\section{Conclusions}\label{sec:Conclusions}
In this work we have modelled the winds of fifteen young, Solar-type stars with well constrained ages in the Pleiades cluster, the AB Doradus moving group, the Columba association and the \(\beta\) Pictoris association. 
These stars aged \SIrange{24}{125}{\mega\year} have a wide range of rotation periods from \SIrange{0.3}{7}{\day}; this is expected for stars in this age range, and matches the observations of e.g.~\citet{2013A&A...556A..36G}.  The models are driven using observationally based magnetic maps derived with Zeeman-Doppler imaging. 

We have studied the coronal magnetic field, Alfvén surface, and wind pressure out to \SI{1}{\astronomicalunit} of our models and we find mostly dipole-dominated coronal magnetic fields; in our models these appear if \SI{\sim 30}{\percent} or more of the surface magnetic energy is dipolar. Our ZDI maps have a wide range of inclinations of the magnetic dipole, and produce a wide range of Alfvén surface lobe and current sheet orientations.
The effective magnetogram degree measures in \(\ell_{.99}\) (see Table~\ref{tab:magnetic_averages}) does not seem to greatly affect the shape of the Alfvén surface in Fig.~\ref{fig:ur-alfven} with the possible exception of the conical Alfvén surfaces of \C1I, \C1M, \C5I, and \C5M. We do not consider these shapes to result from the  \(\ell_{.99}\) values however, but rather from the two stars' very rapid rotational periods which also leads to more detailed ZDI magnetograms as Section~\ref{sec:mag-map-zdi} briefly addressed.

In order to study the isolated effect of the surface magnetic field strength independently of the other model parameters, we created the \(5B_\ZDI\) model series where all magnetic field maps are scaled by a factor of 5. 
Such an increase of the magnetic field strength is generally accompanied by an increase in the mass- and angular momentum loss rate and wind pressure for an Earth-like planet as was seen in Section~\ref{sec:magnetic-scaling}. 
The comparison of the wind models in the \(B_\ZDI\) and \(5B_\ZDI\) series, including the models originally presented in~\citet{paper1,paper2} and re-processed as part of this work (see Section~\ref{sec:numerical_model}),  showed that for this wide range of surface magnetic field strengths there is no single scaling law of the form \(y \propto |B_r|^\alpha\) or \(y \propto \Phi^\alpha\) that is a good fit to the data in the left panel of Fig.~\ref{fig:orthogonal-trend-Unsigned-radial-flux-at-surface}. Instead, as can be seen in the right panel of Fig.~\ref{fig:orthogonal-trend-Unsigned-radial-flux-at-surface}, the  power-law index \(\alpha\) appears to decrease with increasing magnetic field strength.

In Section~\ref{sec:statistical-trends-and-correlations} we fitted power-laws and prediction bands to the \(B_\ZDI\) and \(5B_\ZDI\) model series separately. 
This kind of fitted curve differs from the barbell curves in Section~\ref{sec:magnetic-scaling} as it includes the effect of other stellar parameter that are known to vary with the surface magnetic field, most notably the stellar rotation rate \(\Omega\). 
We find somewhat wider prediction bands than in~\citet{paper2}, suggesting that the more rapid rotation of the stars modelled in this work yield slightly different characteristics than the older stars of~\citet{paper1,paper2}; this can be due to either \(\Omega\) itself or the influence of \(\Omega\) on the effective magnetogram degree. 
In general we find tighter correlations between the unsigned surface flux \(\Phi=4\pi R^2 |B_r|\) and our parameters of interest than between \(|B_r|\) and our parameters of interest. 
This is expected as \(\Phi\) accommodates stellar radius variations (from stellar type, see Table~\ref{tab:observed_quantities}) on the prediction band widths.
The \(\dot J / \Omega\) parameter has a significantly tighter correlation with the magnetic field strength than the \(\dot J\) parameter itself, which is expected from the dependence of \(\dot J\) on \(\Omega\) in eq.~\eqref{eq:angmom_loss}.

In Section~\ref{sec:comparison_with_literture_values} we observe that our results agree well with other models using the \awsom{} model, but predict lower values of \(\dot M\) and \(\dot J\) than models using polytropic MHD models. Only for the strongest magnetic fields in our \(5B_\ZDI\) model series do we obtain comparable values of \(\dot M\) and \(\dot J\) to polytropic MHD models with significantly lower surface magnetic fields. As was previously noted in~\citet{paper2} this gap may possibly be closed by applying an age- or rotation-based scaling to \awsom{} model parameters such as the Poynting flux-to-field ratio \(\Pi_\Alfven/B\), 
which was found by~\citet{2020A&A...635A.178B} to affect the mass loss rate such that \(\dot M \propto \Pi_\Alfven / B\) \citep[see also][]{2021MNRAS.504.1511K}.

In this work and~\citet{paper1,paper2} we have created what we believe to be the largest set of ZDI-driven three-dimensional stellar wind models to date. Between the models all physical and numerical parameters are the same except the surface magnetic maps, and three scalar stellar parameters mass, radius, and rotation rate. This consistency has enabled us to formulate robust scaling relations that includes uncertainty estimates and give predictive ranges for values such as wind mass- and angular momentum loss rates for stars in the \SIrange{0.02}{0.7}{\giga\year} range. We hope that this dataset and our derived scaling relations will be useful in comparing numerical wind models with observational constraints on \(\dot M\), for creating more detailed models of wind-planet interactions, and to give better constraints on how Solar parameters values are applicable to young, Solar-type stars in magnetohydrodynamic modelling.

\section*{Acknowledgements}
    For part of this work, DE has been funded by a University of Southern Queensland 
    (UniSQ) International Stipend Research Scholarship (ISRS) and a
    UniSQ International Fees Research Scholarship (IFRS). 
    Parts of this project has received funding from the European Research Council (ERC) under the European Union's Horizon 2020 research and innovation programme (grant agreement No 817540, ASTROFLOW). 
    Parts of this material is based upon work supported by the National Science Foundation under Grant No. (AST-2108985).
    Parts of this research was undertaken using the University of Southern Queensland Fawkes HPC which is co-sponsored by the Queensland Cyber Infrastructure Foundation (QCIF), see~\url{www.unisq.edu.au/hpc}. 
    We thank SURF (\url{www.surf.nl}) for the support in using the National Supercomputer Snellius.
    This work utilises data obtained by the \href{https://gong.nso.edu/}{Global Oscillation Network Group (GONG)} Program, managed by the National Solar Observatory, which is operated by AURA, Inc.\ under a cooperative agreement with the National Science Foundation. The data were acquired by instruments operated by the Big Bear Solar Observatory, High Altitude Observatory, Learmonth Solar Observatory, Udaipur Solar Observatory,
    Instituto de Astrofísica de Canarias, and Cerro Tololo 
    Interamerican Observatory.
    This work has made use of the \href{http://vald.astro.uu.se/}{Vienna Atomic Line Database (VALD)}, operated at Uppsala University, the Institute of Astronomy RAS in Moscow, and the University of Vienna.
    We acknowledge the use of the \href{http://simbad.u-strasbg.fr/simbad/}{SIMBAD database}. 
    This research has made use of the VizieR catalogue access tool, CDS, Strasbourg, France (\href{https://www.doi.org/10.26093/cds/vizier}{DOI:\@ 10.26093/cds/vizier}). The original description of the VizieR service was published in~\defcitealias{2000A&AS..143...23O}{A\&AS~143,~23}\citetalias{2000A&AS..143...23O}.
    This research has made use of NASA's \href{https://ui.adsabs.harvard.edu/}{Astrophysics Data System}.
    This work was carried out using the \swmf{} tools developed at The University of Michigan \href{https://spaceweather.engin.umich.edu/the-center-for-space-environment-modelling-csem/}{Center for Space Environment Modelling (CSEM)} and made available through the NASA \href{https://ccmc.gsfc.nasa.gov/}{Community Coordinated Modelling Center (CCMC)}. The open source version of \swmf{} is available at 
    \url{https://github.com/MSTEM-QUDA/SWMF.git}. Version \href{https://github.com/MSTEM-QUDA/SWMF/commit/a5beb110f}{2.40.{\tt a5beb110f}} (2022-10-08) has been used for all processing.
    This work has made use of the following additional numerical software, statistics software and visualisation software: 
          NumPy~\citep{2011CSE....13b..22V} version 1.23.5,
          SciPy~\citep{2020SciPy-NMeth} version 1.9.3,
     Matplotlib~\citep{2007CSE.....9...90H} version 3.7.1, 
    statsmodels~\citep{seabold2010statsmodels} version 0.13.2,
    \href{https://www.tecplot.com}{Tecplot} version 2022.1.0.14449, and 
    \href{https://pypi.org/project/pytecplot/}{PyTecplot} version 1.5.0.
    We would like to express our gratitude to the anonymous reviewers for their valuable contributions, which have enhanced the quality and rigour of our article.

\section*{Data availability}
The data underlying this article will be shared on reasonable request to the corresponding author.

%% file: tables/table1.tex
\newcommand{\FT}{\textsuperscript{\dag}}  %
\begin{tabular}{
llll
S[table-format=4.0(2)]  %
S[table-format=2.2(2)]  %
S[table-format=3.0(2)]  %
S[table-format=1.2(2)]  %
S[table-format=1.2(2)]  %
l %
}
\toprule
{Case name}  & 
{Full name \citepalias[see][]{2016MNRAS.457..580F,2018MNRAS.474.4956F}}  & 
{Assoc.}  & 
{Type}  & 
{\(T_\text{eff}\)}  & 
{\(P_\text{rot}\) } &  
{Age  } &  
{\(R\)}  & 
{\(M\)}  & 
{Ref.} \\
& & & & {(\si{\kelvin})} & {(d)} &  {(Myr)}  &  {\((R_\odot)\)} & {\((M_\odot)\)} & \\
\midrule
\Unscaled{F}{C3} HII~296    & \href{http://simbad.u-strasbg.fr/simbad/sim-id?Ident=HII+296                 }{HII 296                }                 & Pleiades           & G8    & 5322 \pm 101 &  2.61 \pm 0.00 & 125 \pm  8 & 0.94 \pm 0.05 & 0.90 \pm 0.04 & \citetalias{2016MNRAS.457..580F} \\ 
\Unscaled{G}{C3} HII~739    & \href{http://simbad.u-strasbg.fr/simbad/sim-id?Ident=HII+739                 }{HII 739                }                 & Pleiades           & G0    & 6066 \pm  89 &  1.56 \pm 0.01 & 125 \pm  8 & 1.54 \pm 0.09 & 1.15 \pm 0.06 & \citetalias{2016MNRAS.457..580F} \\ 
\Unscaled{H}{C3} PELS~31    & \href{http://simbad.u-strasbg.fr/simbad/sim-id?Ident=Cl*+Melotte+22+P        }{Cl* Melotte 22 PELS 31 }                 & Pleiades           & K2\FT & 5046 \pm 108 &  2.50 \pm 0.10 & 125 \pm  8 & 1.05 \pm 0.05 & 0.95 \pm 0.05 & \citetalias{2016MNRAS.457..580F} \\ 
\midrule
\Unscaled{I}{C4} BD-07~2388 & \href{http://simbad.u-strasbg.fr/simbad/sim-id?Ident=BD-07+2388              }{BD-07 2388             }                 & AB Doradus         & K0    & 5121 \pm 137 &  0.33 \pm 0.00 & 120 \pm 10 & 0.78 \pm 0.09 & 0.85 \pm 0.05 & \citetalias{2018MNRAS.474.4956F} \\ 
\Unscaled{J}{C4} HD~6569    & \href{http://simbad.u-strasbg.fr/simbad/sim-id?Ident=HD+6569                 }{HD 6569                }                 & AB Doradus         & K1    & 5118 \pm  95 &  7.13 \pm 0.05 & 120 \pm 10 & 0.76 \pm 0.03 & 0.85 \pm 0.04 & \citetalias{2018MNRAS.474.4956F} \\ 
\Unscaled{K}{C4} HIP~10272  & \href{http://simbad.u-strasbg.fr/simbad/sim-id?Ident=HIP+10272               }{HIP 10272              }                 & AB Doradus         & K1    & 5281 \pm  79 &  6.13 \pm 0.03 & 120 \pm 10 & 0.80 \pm 0.08 & 0.90 \pm 0.04 & \citetalias{2018MNRAS.474.4956F} \\ 
\Unscaled{L}{C4} HIP~76768  & \href{http://simbad.u-strasbg.fr/simbad/sim-id?Ident=HIP+76768               }{HIP 76768              }                 & AB Doradus         & K5    & 4506 \pm 153 &  3.70 \pm 0.02 & 120 \pm 10 & 0.85 \pm 0.09 & 0.80 \pm 0.07 & \citetalias{2016MNRAS.457..580F} \\ 
\Unscaled{M}{C4} LO~Peg     & \href{http://simbad.u-strasbg.fr/simbad/sim-id?Ident=LO+Peg                  }{LO Peg                 } (HIP 106231)    & AB Doradus         & K3    & 4739 \pm 138 &  0.42 \pm 0.00 & 120 \pm 10 & 0.66 \pm 0.08 & 0.75 \pm 0.04 & \citetalias{2016MNRAS.457..580F} \\ 
\Unscaled{N}{C4} PW~And     & \href{http://simbad.u-strasbg.fr/simbad/sim-id?Ident=PW+And                  }{PW And                 }                 & AB Doradus         & K2    & 5012 \pm 108 &  1.76 \pm 0.00 & 120 \pm 10 & 0.78 \pm 0.04 & 0.85 \pm 0.05 & \citetalias{2016MNRAS.457..580F} \\ 
\Unscaled{O}{C4} TYC~0486   & \href{http://simbad.u-strasbg.fr/simbad/sim-id?Ident=TYC+0486-4943-1         }{TYC 0486-4943-1        }                 & AB Doradus         & K4\FT & 4706 \pm 161 &  3.75 \pm 0.30 & 120 \pm 10 & 0.70 \pm 0.14 & 0.77 \pm 0.04 & \citetalias{2016MNRAS.457..580F} \\ 
\Unscaled{P}{C4} TYC~5164   & \href{http://simbad.u-strasbg.fr/simbad/sim-id?Ident=TYC+5164-567-1          }{TYC 5164-567-1         }                 & AB Doradus         & K1\FT & 5130 \pm 161 &  4.68 \pm 0.06 & 120 \pm 10 & 0.90 \pm 0.18 & 0.90 \pm 0.10 & \citetalias{2016MNRAS.457..580F} \\ 
\midrule
\Unscaled{Q}{C5} BD-16~351  & \href{http://simbad.u-strasbg.fr/simbad/sim-id?Ident=BD-16+351               }{BD-16 351              }                 & Col-Hor-Tuc        & K1\FT & 5243 \pm 105 &  3.21 \pm 0.01 &  42 \pm  6 & 0.96 \pm 0.19 & 0.90 \pm 0.09 & \citetalias{2016MNRAS.457..580F} \\ 
\midrule
\Unscaled{R}{C6} HIP~12545  & \href{http://simbad.u-strasbg.fr/simbad/sim-id?Ident=HIP+12545               }{HIP 12545              }                 & \(\beta\) Pictoris & K6    & 4447 \pm 130 &  4.83 \pm 0.01 &  24 \pm  3 & 0.76 \pm 0.05 & 0.95 \pm 0.05 & \citetalias{2016MNRAS.457..580F} \\ 
\Unscaled{S}{C6} TYC~6349   & \href{http://simbad.u-strasbg.fr/simbad/sim-id?Ident=TYC+6349-0200-1         }{TYC 6349-0200-1        }                 & \(\beta\) Pictoris & K7    & 4359 \pm 131 &  3.41 \pm 0.05 &  24 \pm  3 & 0.93 \pm 0.04 & 0.85 \pm 0.05 & \citetalias{2016MNRAS.457..580F} \\ 
\Unscaled{T}{C6} TYC~6878   & \href{http://simbad.u-strasbg.fr/simbad/sim-id?Ident=TYC+6878-0195-1         }{TYC 6878-0195-1        }                 & \(\beta\) Pictoris & K4    & 4667 \pm 120 &  5.70 \pm 0.06 &  24 \pm  3 & 1.39 \pm 0.28 & 1.15 \pm 0.15 & \citetalias{2016MNRAS.457..580F} \\ 
\bottomrule
\end{tabular}

%% file: tables/aggregate-magnetic-quantities.tex
\begin{tabular}{
        l  %
        S[table-format=3.1]  %
        S[table-format=4.1]  %
        S[table-format=3.1]  %
        S[table-format=1.1e2] %
        S[table-format=2.1]  %
        S[table-format=2.1]  %
        S[table-format=2.1]  %
        S[table-format=2.1]  %
        S[table-format=2.0, round-precision=0]  %
        S[table-format=2.0, round-precision=0]  %
}
    \toprule
    Case & {\(|B_r|\)} & {\(\max |{B_r}|\)}  & {\(|\vec B|\)} & {\(\SF\)} & {Dip.} & {Quad.} & {Oct.} & {16+} & {\(\ell_{.90}\)} & {\(\ell_{.99}\)}\\
     & {\((\si{\gauss}) \)}
     & {\((\si{\gauss}) \)}
     & {\((\si{\gauss}) \)}
     & {\((\si{\weber}) \)}
     & {\((\si{\percent}) \)}
     & {\((\si{\percent}) \)}
     & {\((\si{\percent}) \)}
     & {\((\si{\percent}) \)} \\
\midrule
\SimCase{1}{F} & 51.4017    & 221.8806   & 73.87      & 2.75e+16   & 68.67      & 8.35       & 9.53       & 13.45      & 4          & 9         \\
\SimCase{1}{G} & 9.9206     & 45.1025    & 14.35      & 1.43e+16   & 24.73      & 14.78      & 20.43      & 40.06      & 6          & 9         \\
\SimCase{1}{H} & 17.0272    & 105.0705   & 25.13      & 1.14e+16   & 10.43      & 8.96       & 20.39      & 60.23      & 6          & 8         \\
\midrule
\SimCase{1}{I} & 94.1571    & 536.3951   & 136.92     & 3.51e+16   & 43.75      & 4.52       & 3.71       & 48.01      & 12         & 15        \\
\SimCase{1}{J} & 15.2727    & 41.6936    & 21.45      & 5.35e+15   & 76.32      & 19.49      & 3.17       & 1.02       & 2          & 4         \\
\SimCase{1}{K} & 10.0653    & 29.8653    & 13.76      & 3.93e+15   & 82.56      & 11.19      & 2.49       & 3.76       & 2          & 5         \\
\SimCase{1}{L} & 58.6443    & 222.0428   & 82.54      & 2.60e+16   & 83.73      & 1.81       & 3.62       & 10.83      & 4          & 7         \\
\SimCase{1}{M} & 88.5832    & 588.5568   & 131.28     & 2.38e+16   & 35.98      & 8.35       & 7.41       & 48.27      & 10         & 15        \\
\SimCase{1}{N} & 70.6653    & 448.3880   & 100.83     & 2.65e+16   & 43.51      & 7.59       & 5.66       & 43.23      & 8          & 12        \\
\SimCase{1}{O} & 18.7183    & 59.3175    & 27.66      & 5.58e+15   & 38.12      & 29.97      & 21.25      & 10.66      & 4          & 6         \\
\SimCase{1}{P} & 48.4955    & 128.3140   & 67.94      & 2.38e+16   & 85.61      & 4.50       & 2.58       & 7.31       & 2          & 6         \\
\midrule
\SimCase{1}{Q} & 32.7697    & 163.6843   & 47.47      & 1.86e+16   & 57.76      & 21.42      & 7.95       & 12.88      & 4          & 6         \\
\midrule
\SimCase{1}{R} & 33.4445    & 135.5011   & 47.31      & 1.77e+16   & 56.33      & 7.55       & 8.64       & 27.48      & 6          & 10        \\
\SimCase{1}{S} & 30.0614    & 137.0246   & 43.46      & 3.54e+16   & 61.04      & 8.39       & 10.70      & 19.87      & 5          & 6         \\
\SimCase{1}{T} & 64.9438    & 403.9809   & 94.13      & 2.28e+16   & 53.26      & 12.59      & 13.63      & 20.52      & 4          & 7         \\
\midrule
\SimCase{5}{F} & 256.9816   & 1109.3581  & 370.19     & 1.37e+17   & 68.76      & 8.33       & 9.50       & 13.41      & 4          & 9         \\
\SimCase{5}{G} & 49.5965    & 225.4990   & 71.95      & 7.14e+16   & 25.02      & 14.74      & 20.38      & 39.86      & 6          & 9         \\
\SimCase{5}{H} & 85.1272    & 525.3495   & 125.95     & 5.68e+16   & 10.56      & 8.98       & 20.36      & 60.10      & 6          & 8         \\
\midrule
\SimCase{5}{I} & 470.7491   & 2681.8895  & 685.58     & 1.75e+17   & 43.85      & 4.52       & 3.71       & 47.92      & 12         & 15        \\
\SimCase{5}{J} & 76.3447    & 208.4290   & 107.66     & 2.67e+16   & 76.43      & 19.41      & 3.15       & 1.01       & 2          & 4         \\
\SimCase{5}{K} & 50.3109    & 149.3035   & 69.20      & 1.96e+16   & 82.73      & 11.08      & 2.47       & 3.72       & 2          & 5         \\
\SimCase{5}{L} & 293.1898   & 1110.1755  & 413.60     & 1.30e+17   & 83.78      & 1.81       & 3.61       & 10.80      & 4          & 7         \\
\SimCase{5}{M} & 442.8857   & 2942.7782  & 657.24     & 1.19e+17   & 36.06      & 8.34       & 7.40       & 48.20      & 10         & 15        \\
\SimCase{5}{N} & 353.2914   & 2241.8846  & 504.85     & 1.32e+17   & 43.61      & 7.58       & 5.66       & 43.15      & 8          & 12        \\
\SimCase{5}{O} & 93.5759    & 296.5540   & 138.69     & 2.79e+16   & 38.34      & 29.85      & 21.17      & 10.63      & 4          & 6         \\
\SimCase{5}{P} & 242.4464   & 641.5275   & 340.55     & 1.19e+17   & 85.66      & 4.49       & 2.57       & 7.28       & 2          & 6         \\
\midrule
\SimCase{5}{Q} & 163.8249   & 818.3774   & 238.15     & 9.28e+16   & 57.90      & 21.36      & 7.92       & 12.83      & 4          & 6         \\
\midrule
\SimCase{5}{R} & 167.2020   & 677.4658   & 237.25     & 8.85e+16   & 56.50      & 7.53       & 8.61       & 27.36      & 6          & 10        \\
\SimCase{5}{S} & 150.2914   & 685.0870   & 218.04     & 1.77e+17   & 61.20      & 8.35       & 10.66      & 19.79      & 5          & 6         \\
\SimCase{5}{T} & 324.6864   & 2019.8525  & 471.34     & 1.14e+17   & 53.34      & 12.57      & 13.61      & 20.48      & 4          & 7         \\
\bottomrule 
\end{tabular}

%% file: tables/main-table-body.tex
\begin{tabular}{
    l
    S[table-format=1.2, round-precision=2]  %
    S[table-format=1.2, round-precision=2]  %
    S[table-format=2.1, round-precision=1]  %
    S[table-format=1.2, round-precision=2]  %
    S[table-format=2.1, round-precision=1]  %
    S[table-format=2.1, round-precision=1]  %
    S[table-format=1.1e3, round-precision=1]  %
    S[table-format=1.1e3, round-precision=1]  %
    S[table-format=1.1e3, round-precision=1]  %
    S[table-format=1.1,  round-precision=1]  %
}

\toprule
Case 
& {\(\Phi_\text{open}\)}
& {\(S_\text{open}\)}
& {\(i_{B}\)}
& {\(\Phi_\text{axi}\)}
& {\(R_\Alfven\) }
& {\(|\vec r_\Alfven\times \uvec \Omega|\)}
& {\(\dot M\) }
& {\(\dot J\) }
& {\(P^\Earth_\Wind\)}
& {\(R_\text{m}\) }
\\ %
& {\(\left(\SF\right)\)}
& {\(\left(S_\Star\right)\)}
& {\(\left(\si{\degree}\right)\)}
& {\(\left(\Phi_\text{open}\right)\)}
& {\(\left(R_\Star{}\right)\)}
& {\(\left(R_\Star{}\right)\)}
& {\(\left(\si{\kilogram\per\second}\right)\)}
& {\(\left(\si{\newton\meter}\right)\)}
& {\(\left(\si{\pascal}\right)\)}
& {\(\left(R_\text{p}\right)\)}
\\ 
\midrule
\SimCase{1}{F} & 0.22       & 0.10       & 13.6       & 0.92       & 22.5       & 16.8       & 3.1e+10    & 4.3e+25    & 1.0e-07    & 5.2       \\
\SimCase{1}{G} & 0.18       & 0.09       & 51.9       & 0.63       & 12.1       & 9.6        & 2.2e+10    & 9.3e+25    & 5.4e-08    & 5.8       \\
\SimCase{1}{H} & 0.16       & 0.12       & 43.5       & 0.67       & 11.8       & 9.1        & 1.7e+10    & 1.7e+25    & 2.4e-08    & 6.6       \\
\midrule
\SimCase{1}{I} & 0.17       & 0.09       & 8.3        & 0.99       & 30.2       & 22.7       & 3.5e+10    & 3.9e+26    & 4.4e-07    & 4.1       \\
\SimCase{1}{J} & 0.30       & 0.16       & 32.4       & 0.71       & 16.1       & 12.3       & 7.4e+09    & 2.1e+24    & 1.2e-08    & 7.4       \\
\SimCase{1}{K} & 0.32       & 0.21       & 49.8       & 0.51       & 14.5       & 11.3       & 5.1e+09    & 1.9e+24    & 6.5e-09    & 8.2       \\
\SimCase{1}{L} & 0.22       & 0.11       & 7.5        & 0.97       & 23.1       & 17.2       & 3.0e+10    & 2.3e+25    & 1.3e-07    & 4.9       \\
\SimCase{1}{M} & 0.16       & 0.09       & 11.3       & 0.97       & 29.0       & 22.1       & 2.2e+10    & 1.3e+26    & 1.6e-07    & 4.8       \\
\SimCase{1}{N} & 0.17       & 0.12       & 33.1       & 0.76       & 24.7       & 19.0       & 2.5e+10    & 5.5e+25    & 5.1e-08    & 5.8       \\
\SimCase{1}{O} & 0.24       & 0.11       & 87.5       & 0.08       & 14.9       & 11.8       & 8.6e+09    & 4.6e+24    & 7.2e-09    & 8.0       \\
\SimCase{1}{P} & 0.23       & 0.13       & 5.7        & 0.98       & 21.6       & 16.1       & 3.0e+10    & 1.9e+25    & 1.2e-07    & 5.0       \\
\midrule
\SimCase{1}{Q} & 0.24       & 0.13       & 71.2       & 0.27       & 20.5       & 16.3       & 2.4e+10    & 4.4e+25    & 2.3e-08    & 6.6       \\
\midrule
\SimCase{1}{R} & 0.23       & 0.16       & 34.8       & 0.70       & 19.9       & 15.2       & 2.2e+10    & 2.7e+25    & 2.7e-08    & 6.5       \\
\SimCase{1}{S} & 0.25       & 0.13       & 49.2       & 0.53       & 19.8       & 15.4       & 4.0e+10    & 7.2e+25    & 5.3e-08    & 5.8       \\
\SimCase{1}{T} & 0.19       & 0.11       & 19.7       & 0.86       & 23.9       & 18.0       & 2.2e+10    & 1.2e+25    & 5.9e-08    & 5.7       \\
\midrule
\SimCase{5}{F} & 0.14       & 0.07       & 13.9       & 0.93       & 36.1       & 26.9       & 9.1e+10    & 1.6e+26    & 4.4e-07    & 4.0       \\
\SimCase{5}{G} & 0.13       & 0.08       & 48.2       & 0.71       & 20.8       & 16.7       & 6.9e+10    & 5.3e+26    & 2.4e-07    & 4.5       \\
\SimCase{5}{H} & 0.11       & 0.09       & 40.3       & 0.70       & 18.9       & 14.6       & 5.3e+10    & 9.5e+25    & 8.8e-08    & 5.3       \\
\midrule
\SimCase{5}{I}$^{\dagger}$ & 0.11       & 0.06       & 8.9        & 0.99       & {---}      & {---}      & 7.8e+10    & 1.6e+27    & 9.0e-07    & 3.6       \\
\SimCase{5}{J} & 0.20       & 0.10       & 32.3       & 0.71       & 25.6       & 19.5       & 2.8e+10    & 1.4e+25    & 5.3e-08    & 5.8       \\
\SimCase{5}{K} & 0.22       & 0.14       & 49.5       & 0.51       & 23.0       & 17.8       & 2.2e+10    & 1.6e+25    & 3.0e-08    & 6.4       \\
\SimCase{5}{L} & 0.14       & 0.07       & 7.6        & 0.97       & 36.7       & 27.3       & 9.0e+10    & 8.2e+25    & 4.9e-07    & 4.0       \\
\SimCase{5}{M} & 0.11       & 0.06       & 12.3       & 0.98       & 46.2       & 34.9       & 5.4e+10    & 5.3e+26    & 3.8e-07    & 4.2       \\
\SimCase{5}{N} & 0.11       & 0.08       & 32.8       & 0.80       & 39.3       & 30.1       & 7.1e+10    & 2.6e+26    & 1.6e-07    & 4.8       \\
\SimCase{5}{O} & 0.16       & 0.07       & 87.3       & 0.08       & 23.7       & 18.9       & 2.8e+10    & 3.2e+25    & 3.5e-08    & 6.2       \\
\SimCase{5}{P} & 0.14       & 0.08       & 5.8        & 0.98       & 34.3       & 25.5       & 8.8e+10    & 7.0e+25    & 4.6e-07    & 4.0       \\
\midrule
\SimCase{5}{Q} & 0.15       & 0.08       & 70.0       & 0.31       & 32.9       & 26.0       & 7.2e+10    & 2.8e+26    & 8.9e-08    & 5.3       \\
\midrule
\SimCase{5}{R} & 0.15       & 0.10       & 34.1       & 0.72       & 32.0       & 24.4       & 6.8e+10    & 1.5e+26    & 1.1e-07    & 5.1       \\
\SimCase{5}{S} & 0.16       & 0.08       & 48.5       & 0.55       & 32.2       & 24.9       & 1.3e+11    & 4.3e+26    & 2.0e-07    & 4.6       \\
\SimCase{5}{T} & 0.12       & 0.07       & 20.0       & 0.86       & 39.3       & 29.4       & 6.1e+10    & 5.3e+25    & 2.3e-07    & 4.5       \\
\bottomrule
\end{tabular}

%% file: tables/correlations-rvalues-pvalues-predint-1-5.tex
\begin{tabular}{
    l
    S[table-format=.2(2)]  %
    S[table-format=+1.2(2)]  %
    S[table-format=1.2]     %
    S[table-format=2.2]     %
    S[table-format=1.1e+2]   %
    S[table-format=+1.2(2)]  %
    S[table-format=+1.2(2)]  %
    S[table-format=1.2]     %
    S[table-format=2.2]     %
    S[table-format=1.1e+2]   %
    }
    \toprule
{Quantity} & \multicolumn{10}{c}{Correlation with $a \log_{10}|B_r| + \log_{10} b$} \\
\cmidrule(lr){2-11}
    {} & \multicolumn{5}{c}{$B_\ZDI$ series}& \multicolumn{5}{c}{$5B_\ZDI$ series}\\
    \cmidrule(lr){2-6}\cmidrule(lr){7-11}
    {} & 
    {$a$} & {$b$} & {$r^2$} & {$\frac{y_\text{0.975}}{y_\text{0.025}}$} & {$p$} &
    {$a$} & {$b$} & {$r^2$} & {$\frac{y_\text{0.975}}{y_\text{0.025}}$} & {$p$} \\
    \midrule
$\log_{10}\max|B_r|$                                 &   1.12 \pm 0.11 &  0.46 \pm 0.15 &  0.94 &  2.85 &2.4e-18&   1.12 \pm 0.11 &  0.37 \pm 0.23 &  0.94 &  2.85 &2.4e-18\\
$\log_{10}|\vec{B}|$                                 &   1.01 \pm 0.01 &  0.15 \pm 0.01 & 0.999 &  1.09 &1.9e-46&   1.00 \pm 0.01 &  0.15 \pm 0.02 & 0.999 &  1.09 &9.6e-47\\
$\log_{10}\Phi$                                      &   0.97 \pm 0.16 & 14.69 \pm 0.22 &  0.84 &  4.57 &1.1e-12&   0.97 \pm 0.16 & 14.71 \pm 0.33 &  0.84 &  4.57 &1.1e-12\\
$\log_{10}\Phi_\text{open}$                          &   0.75 \pm 0.16 & 14.37 \pm 0.22 &  0.76 &  4.53 &3.8e-10&   0.71 \pm 0.16 & 14.44 \pm 0.33 &  0.74 &  4.50 &9.5e-10\\
$\log_{10} R_\Alfven$                                &   0.33 \pm 0.04 &  0.79 \pm 0.05 &  0.91 &  1.45 &4.4e-16&   0.34 \pm 0.04 &  0.73 \pm 0.09 &  0.90 &  1.49 &9.3e-16\\
$\log_{10} |\vec{r_\Alfven} \times \uvec{\Omega}|$   &   0.31 \pm 0.04 &  0.70 \pm 0.06 &  0.89 &  1.49 &7.1e-15&   0.32 \pm 0.05 &  0.66 \pm 0.09 &  0.88 &  1.54 &2.1e-14\\
$\log_{10}\dot M$                                    &   0.81 \pm 0.18 &  9.01 \pm 0.24 &  0.76 &  5.25 &4.4e-10&   0.62 \pm 0.16 &  9.35 \pm 0.33 &  0.69 &  4.51 &1.5e-08\\
$\log_{10}\dot J$                                    &   1.85 \pm 0.49 & 22.42 \pm 0.66 &  0.68 & 98.32 &2.4e-08&   1.55 \pm 0.49 & 22.50 \pm 0.99 &  0.60 & 94.02 &4.8e-07\\
$\log_{10}\dot J / \Omega$                           &   1.11 \pm 0.36 & 28.18 \pm 0.48 &  0.58 & 29.35 &9.5e-07&   0.81 \pm 0.36 & 28.77 \pm 0.73 &  0.43 & 28.76 &7.9e-05\\
$\log_{10}P_\text{wind}^\Earth$                      &   1.29 \pm 0.27 & -9.35 \pm 0.37 &  0.77 & 12.85 &2.3e-10&   1.10 \pm 0.26 & -9.27 \pm 0.53 &  0.73 & 11.54 &2.4e-09\\
$\log_{10} R_\text{mag}$                             &  -0.21 \pm 0.05 &  1.11 \pm 0.06 &  0.77 &  1.53 &2.3e-10&  -0.18 \pm 0.04 &  1.09 \pm 0.09 &  0.73 &  1.50 &2.4e-09\\
\midrule
{Quantity} & \multicolumn{10}{c}{Correlation with $a \log_{10}\Phi + \log_{10} b$} \\
\cmidrule(lr){2-11}
    {} & \multicolumn{5}{c}{$B_\ZDI$ series}& \multicolumn{5}{c}{$5B_\ZDI$ series}\\
    \cmidrule(lr){2-6}\cmidrule(lr){7-11}
    {} & 
    {$a$} & {$b$} & {$r^2$} & {$\frac{y_\text{0.975}}{y_\text{0.025}}$} & {$p$} &
    {$a$} & {$b$} & {$r^2$} & {$\frac{y_\text{0.975}}{y_\text{0.025}}$} & {$p$} \\
    \midrule
$\log_{10}|B_r|$                                     &   0.86 \pm 0.15 &-12.50 \pm 2.32 &  0.84 &  4.19 &1.1e-12&   0.87 \pm 0.15 &-12.40 \pm 2.42 &  0.84 &  4.19 &1.1e-12\\
$\log_{10}\max|B_r|$                                 &   0.99 \pm 0.18 &-13.88 \pm 2.90 &  0.82 &  5.99 &8.2e-12&   0.99 \pm 0.18 &-13.87 \pm 3.02 &  0.82 &  5.99 &8.2e-12\\
$\log_{10}|\vec{B}|$                                 &   0.87 \pm 0.15 &-12.42 \pm 2.33 &  0.84 &  4.23 &1.1e-12&   0.87 \pm 0.15 &-12.31 \pm 2.43 &  0.84 &  4.22 &1.1e-12\\
$\log_{10}\Phi_\text{open}$                          &   0.79 \pm 0.07 &  2.80 \pm 1.13 &  0.95 &  2.01 &1.4e-19&   0.76 \pm 0.07 &  3.24 \pm 1.13 &  0.95 &  1.95 &1.2e-19\\
$\log_{10} R_\Alfven$                                &   0.28 \pm 0.06 & -3.21 \pm 1.03 &  0.73 &  1.89 &1.6e-09&   0.30 \pm 0.06 & -3.53 \pm 1.05 &  0.77 &  1.86 &2.2e-10\\
$\log_{10} |\vec{r_\Alfven} \times \uvec{\Omega}|$   &   0.27 \pm 0.06 & -3.14 \pm 1.02 &  0.72 &  1.88 &2.8e-09&   0.28 \pm 0.06 & -3.41 \pm 1.03 &  0.76 &  1.84 &3.7e-10\\
$\log_{10}\dot M$                                    &   0.87 \pm 0.05 & -3.86 \pm 0.76 &  0.98 &  1.59 &1.5e-25&   0.69 \pm 0.06 & -0.90 \pm 0.96 &  0.96 &  1.76 &1.7e-20\\
$\log_{10}\dot J$                                    &   1.98 \pm 0.29 & -6.72 \pm 4.67 &  0.87 & 17.93 &5.0e-14&   1.70 \pm 0.32 & -2.75 \pm 5.25 &  0.81 & 22.37 &1.0e-11\\
$\log_{10}\dot J / \Omega$                           &   1.30 \pm 0.17 &  8.93 \pm 2.75 &  0.89 &  5.46 &3.3e-15&   1.02 \pm 0.22 & 13.38 \pm 3.65 &  0.76 &  8.68 &2.7e-10\\
$\log_{10}P_\text{wind}^\Earth$                      &   1.27 \pm 0.21 &-27.98 \pm 3.41 &  0.84 &  8.22 &1.0e-12&   1.11 \pm 0.20 &-25.58 \pm 3.29 &  0.83 &  7.01 &3.9e-12\\
$\log_{10} R_\text{mag}$                             &  -0.21 \pm 0.04 &  4.21 \pm 0.57 &  0.84 &  1.42 &1.0e-12&  -0.19 \pm 0.03 &  3.81 \pm 0.55 &  0.83 &  1.38 &3.9e-12\\
\midrule
{Quantity} & \multicolumn{10}{c}{Correlation with $a \log_{10}\Phi_\text{open} + \log_{10} b$} \\
\cmidrule(lr){2-11}
    {} & \multicolumn{5}{c}{$B_\ZDI$ series}& \multicolumn{5}{c}{$5B_\ZDI$ series}\\
    \cmidrule(lr){2-6}\cmidrule(lr){7-11}
    {} & 
    {$a$} & {$b$} & {$r^2$} & {$\frac{y_\text{0.975}}{y_\text{0.025}}$} & {$p$} &
    {$a$} & {$b$} & {$r^2$} & {$\frac{y_\text{0.975}}{y_\text{0.025}}$} & {$p$} \\
    \midrule
$\log_{10}|B_r|$                                     &   1.02 \pm 0.22 &-14.32 \pm 3.40 &  0.76 &  5.84 &3.8e-10&   1.04 \pm 0.24 &-14.57 \pm 3.77 &  0.74 &  6.19 &9.5e-10\\
$\log_{10}\max|B_r|$                                 &   1.14 \pm 0.29 &-15.57 \pm 4.37 &  0.71 &  9.65 &6.5e-09&   1.17 \pm 0.30 &-15.98 \pm 4.78 &  0.69 & 10.10 &1.1e-08\\
$\log_{10}|\vec{B}|$                                 &   1.02 \pm 0.23 &-14.20 \pm 3.46 &  0.75 &  6.00 &4.9e-10&   1.05 \pm 0.24 &-14.44 \pm 3.81 &  0.74 &  6.32 &1.2e-09\\
$\log_{10}\Phi$                                      &   1.21 \pm 0.11 & -2.56 \pm 1.67 &  0.95 &  2.37 &1.4e-19&   1.25 \pm 0.11 & -3.20 \pm 1.78 &  0.95 &  2.36 &1.2e-19\\
$\log_{10} R_\Alfven$                                &   0.35 \pm 0.08 & -4.09 \pm 1.20 &  0.74 &  1.87 &8.6e-10&   0.38 \pm 0.08 & -4.64 \pm 1.29 &  0.77 &  1.86 &2.2e-10\\
$\log_{10} |\vec{r_\Alfven} \times \uvec{\Omega}|$   &   0.33 \pm 0.08 & -4.01 \pm 1.18 &  0.74 &  1.84 &1.1e-09&   0.37 \pm 0.08 & -4.50 \pm 1.23 &  0.77 &  1.82 &2.2e-10\\
$\log_{10}\dot M$                                    &   1.06 \pm 0.10 & -6.18 \pm 1.57 &  0.94 &  2.26 &9.2e-19&   0.88 \pm 0.08 & -3.40 \pm 1.29 &  0.95 &  1.87 &2.7e-19\\
$\log_{10}\dot J$                                    &   2.34 \pm 0.46 &-11.11 \pm 7.01 &  0.80 & 37.84 &3.3e-11&   2.12 \pm 0.45 & -8.09 \pm 7.19 &  0.77 & 32.36 &2.4e-10\\
$\log_{10}\dot J / \Omega$                           &   1.63 \pm 0.19 &  4.67 \pm 2.91 &  0.92 &  4.52 &1.2e-16&   1.37 \pm 0.24 &  8.68 \pm 3.79 &  0.83 &  6.23 &2.5e-12\\
$\log_{10}P_\text{wind}^\Earth$                      &   1.51 \pm 0.32 &-30.79 \pm 4.92 &  0.77 & 12.79 &2.2e-10&   1.37 \pm 0.30 &-28.81 \pm 4.72 &  0.76 &  9.80 &3.5e-10\\
$\log_{10} R_\text{mag}$                             &  -0.25 \pm 0.05 &  4.68 \pm 0.82 &  0.77 &  1.53 &2.2e-10&  -0.23 \pm 0.05 &  4.35 \pm 0.79 &  0.76 &  1.46 &3.5e-10\\
\bottomrule
\end{tabular}

%% file: reprocessed-table.tex
\section{Reprocessed aggregate parameters}\label{sec:full-table}
In Section~\ref{sec:Simulations} we noted that the Hyades \citep{paper1} and Coma Berenices and Hercules-Lyra \citep{paper2} models have been re-processed with the same numerical configuration as the models presented in this work. In Table~\ref{tab:main-table-appendix} we provide the aggregate parameters (in the same format as Table~\ref{tab:main-table}) resulting from the re-processing. In the re-processing we found only minor differences to the results in \citet{paper1,paper2} so that the overall conclusions of this work and \citet{paper1,paper2} were not affected by the re-processing.

Note that the Solar maximum/minimum models were based on magnetograms from the Global Oscillation Network Group (GONG) spherical harmonic transform coefficients\footnote{GONG data is available at \url{https://gong.nso.edu/data/magmap/}.} at Carrington Rotation 2157 (maximum) and 2211 (minimum) and are thus referred to as Sun-G2157 and Sun-G2211 in \citetalias{paper1}.

\begin{table*}
    \centering
    \caption{
        Aggregate parameters calculated from the re-processed wind models of \citetalias{paper1,paper2}. The row structure of this table is identical to that of Table~\ref{tab:main-table}, except that the rightmost column provides a reference to the paper where this model was originally published. For further information about the solar and stellar models listed here we refer the reader to \citetalias{paper1,paper2}. 
    }\label{tab:main-table-appendix}
    \sisetup{
        table-figures-decimal=1,
        table-figures-integer=1,
        table-figures-uncertainty=0,
        table-figures-exponent = 3,
        table-number-alignment=center,
        round-mode=places,
        round-precision=1
    }
    \input{tables/main-table-appendix.tex}

\end{table*}

%% file: tables/main-table-appendix.tex
\begin{tabular}{
    l
    S[table-format=1.2, round-precision=2]  %
    S[table-format=1.2, round-precision=2]  %
    S[table-format=2.1, round-precision=1]  %
    S[table-format=1.2, round-precision=2]  %
    S[table-format=2.1, round-precision=1]  %
    S[table-format=2.1, round-precision=1]  %
    S[table-format=1.1e3, round-precision=1]  %
    S[table-format=1.1e3, round-precision=1]  %
    S[table-format=1.1e3, round-precision=1]  %
    S[table-format=1.1,  round-precision=1]  %
    l
}

\toprule
Case 
& {\(\Phi_\text{open}\)}
& {\(S_\text{open}\)}
& {\(i_{B_r=0}\)}
& {\(\Phi_\text{axi}\)}
& {\(R_\Alfven\) }
& {\(|\vec r_\Alfven\times \uvec \Omega|\)}
& {\(\dot M\) }
& {\(\dot J\) }
& {\(P^\Earth_\Wind\)}
& {\(R_\text{m}\) }
& Orig.
\\ %
& {\(\left(\SF\right)\)}
& {\(\left(S\right)\)}
& {\(\left(\si{\degree}\right)\)}
& {\(\left(\Phi_\text{open}\right)\)}
& {\(\left(R_\Star{}\right)\)}
& {\(\left(R_\Star{}\right)\)}
& {\(\left(\si{\kilogram\per\second}\right)\)}
& {\(\left(\si{\newton\meter}\right)\)}
& {\(\left(\si{\pascal}\right)\)}
& {\(\left(R_\text{p}\right)\)}
& paper
\\ 
\midrule
\(\odot\) Sun (maximum) & 0.12       & 0.05       & 66.9       & 0.38       & 6.9        & 5.4        & 3.8e+09    & 6.8e+22    & 4.2e-09    & 8.8       &\citetalias{paper1}\\
\(\odot\) Sun (minimum) & 0.38       & 0.23       & 21.0       & 0.86       & 5.0        & 3.9        & 3.9e+08    & 3.8e+21    & 6.2e-10    & 12.1      &\citetalias{paper1}\\
\midrule
\SimSymbol{1}{0} Mel25-5 & 0.38       & 0.22       & 67.0       & 0.31       & 12.0       & 9.4        & 4.9e+09    & 9.8e+23    & 5.7e-09    & 8.4       &\citetalias{paper1}\\
\SimSymbol{1}{1} Mel25-21 & 0.35       & 0.21       & 55.9       & 0.43       & 14.4       & 11.2       & 7.5e+09    & 2.1e+24    & 9.3e-09    & 7.7      &\citetalias{paper1}\\
\SimSymbol{1}{2} Mel25-43 & 0.39       & 0.23       & 89.0       & 0.02       & 12.4       & 9.9        & 3.3e+09    & 5.8e+23    & 3.9e-09    & 8.9      &\citetalias{paper1}\\
\SimSymbol{1}{3} Mel25-151 & 0.30       & 0.18       & 53.7       & 0.46       & 13.8       & 10.7       & 6.6e+09    & 1.3e+24    & 8.1e-09    & 7.9     &\citetalias{paper1}\\
\SimSymbol{1}{4} Mel25-179 & 0.29       & 0.17       & 46.8       & 0.54       & 16.2       & 12.5       & 1.0e+10    & 3.0e+24    & 1.4e-08    & 7.2     &\citetalias{paper1}\\
\midrule
\SimSymbol{1}{5} AV 523 & 0.26       & 0.23       & 3.5        & 0.99       & 12.9       & 9.7        & 4.1e+09    & 3.8e+23    & 1.1e-08    & 7.5       &\citetalias{paper2}\\
\SimSymbol{1}{6} AV 1693 & 0.22       & 0.20       & 46.3       & 0.56       & 14.8       & 11.4       & 1.2e+10    & 2.8e+24    & 1.5e-08    & 7.1       &\citetalias{paper2}\\
\SimSymbol{1}{7} AV 1826 & 0.25       & 0.16       & 15.8       & 0.88       & 12.7       & 9.6        & 7.3e+09    & 1.0e+24    & 1.3e-08    & 7.3       &\citetalias{paper2}\\
\SimSymbol{1}{8} AV 2177 & 0.36       & 0.15       & 85.2       & 0.09       & 11.7       & 9.3        & 2.8e+09    & 4.9e+23    & 3.1e-09    & 9.2       &\citetalias{paper2}\\
\SimSymbol{1}{9} TYC 1987 & 0.25       & 0.17       & 30.6       & 0.74       & 14.0       & 10.6       & 8.3e+09    & 1.6e+24    & 1.5e-08    & 7.1       &\citetalias{paper2}\\
\midrule
\SimSymbol{1}{A} DX Leo & 0.28       & 0.17       & 82.5       & 0.11       & 17.8       & 14.3       & 1.1e+10    & 7.6e+24    & 1.3e-08    & 7.3       &\citetalias{paper2}\\
\SimSymbol{1}{B} EP Eri & 0.25       & 0.14       & 76.1       & 0.32       & 11.7       & 9.2        & 4.0e+09    & 7.6e+23    & 4.1e-09    & 8.8       &\citetalias{paper2}\\
\SimSymbol{1}{C} HH Leo & 0.29       & 0.13       & 82.3       & 0.11       & 15.8       & 12.6       & 9.3e+09    & 4.9e+24    & 1.1e-08    & 7.5       &\citetalias{paper2}\\
\SimSymbol{1}{D} V439 And & 0.34       & 0.20       & 8.8        & 0.96       & 13.4       & 10.1       & 6.8e+09    & 2.1e+24    & 2.0e-08    & 6.8       &\citetalias{paper2}\\
\SimSymbol{1}{E} V447 Lac & 0.25       & 0.23       & 25.4       & 0.81       & 12.6       & 9.5        & 5.8e+09    & 1.8e+24    & 8.7e-09    & 7.8       &\citetalias{paper2}\\
\midrule
\SimSymbol{5}{0} Mel25-5 & 0.26       & 0.14       & 66.2       & 0.30       & 18.7       & 14.8       & 2.0e+10    & 8.3e+24    & 2.4e-08    & 6.6       &\citetalias{paper1}\\
\SimSymbol{5}{1} Mel25-21 & 0.23       & 0.14       & 56.0       & 0.43       & 22.7       & 17.7       & 2.8e+10    & 1.6e+25    & 3.8e-08    & 6.1       &\citetalias{paper1}\\
\SimSymbol{5}{2} Mel25-43 & 0.27       & 0.15       & 89.0       & 0.02       & 19.2       & 15.4       & 1.5e+10    & 5.6e+24    & 1.8e-08    & 6.9       &\citetalias{paper1}\\
\SimSymbol{5}{3} Mel25-151 & 0.20       & 0.12       & 53.9       & 0.44       & 21.9       & 17.0       & 2.3e+10    & 9.1e+24    & 3.2e-08    & 6.3       &\citetalias{paper1}\\
\SimSymbol{5}{4} Mel25-179 & 0.19       & 0.11       & 46.7       & 0.53       & 25.9       & 20.0       & 3.6e+10    & 2.0e+25    & 5.3e-08    & 5.8       &\citetalias{paper1}\\
\midrule
\SimSymbol{5}{5} AV 523 & 0.18       & 0.16       & 3.5        & 0.99       & 19.5       & 14.5       & 1.7e+10    & 2.7e+24    & 5.8e-08    & 5.7       &\citetalias{paper2}\\
\SimSymbol{5}{6} AV 1693 & 0.15       & 0.13       & 45.6       & 0.55       & 22.8       & 17.6       & 4.1e+10    & 1.9e+25    & 5.9e-08    & 5.7       &\citetalias{paper2}\\
\SimSymbol{5}{7} AV 1826 & 0.16       & 0.10       & 15.7       & 0.88       & 19.4       & 14.5       & 2.7e+10    & 6.1e+24    & 6.4e-08    & 5.6       &\citetalias{paper2}\\
\SimSymbol{5}{8} AV 2177 & 0.25       & 0.10       & 84.8       & 0.09       & 18.3       & 14.6       & 1.3e+10    & 4.7e+24    & 1.6e-08    & 7.1       &\citetalias{paper2}\\
\SimSymbol{5}{9} TYC 1987 & 0.16       & 0.12       & 30.8       & 0.72       & 21.9       & 16.6       & 2.9e+10    & 1.0e+25    & 6.2e-08    & 5.6       &\citetalias{paper2}\\
\midrule
\SimSymbol{5}{A} DX Leo & 0.18       & 0.11       & 81.9       & 0.12       & 28.0       & 22.4       & 4.0e+10    & 5.6e+25    & 5.4e-08    & 5.8       &\citetalias{paper2}\\
\SimSymbol{5}{B} EP Eri & 0.17       & 0.10       & 73.8       & 0.29       & 17.2       & 13.6       & 1.8e+10    & 6.4e+24    & 2.3e-08    & 6.6       &\citetalias{paper2}\\
\SimSymbol{5}{C} HH Leo & 0.19       & 0.08       & 81.8       & 0.12       & 24.9       & 20.0       & 3.2e+10    & 3.5e+25    & 4.3e-08    & 6.0       &\citetalias{paper2}\\
\SimSymbol{5}{D} V439 And & 0.23       & 0.14       & 9.0        & 0.96       & 21.0       & 15.7       & 2.7e+10    & 1.3e+25    & 9.9e-08    & 5.2       &\citetalias{paper2}\\
\SimSymbol{5}{E} V447 Lac & 0.17       & 0.17       & 26.4       & 0.79       & 19.2       & 14.5       & 2.2e+10    & 1.2e+25    & 3.9e-08    & 6.1       &\citetalias{paper2}\\
\bottomrule
\end{tabular}

%% file: pooled-series.tex
\section{Pooled series}\label{sec:pooled-series}
\begin{figure*}
    \centering
    \includegraphics{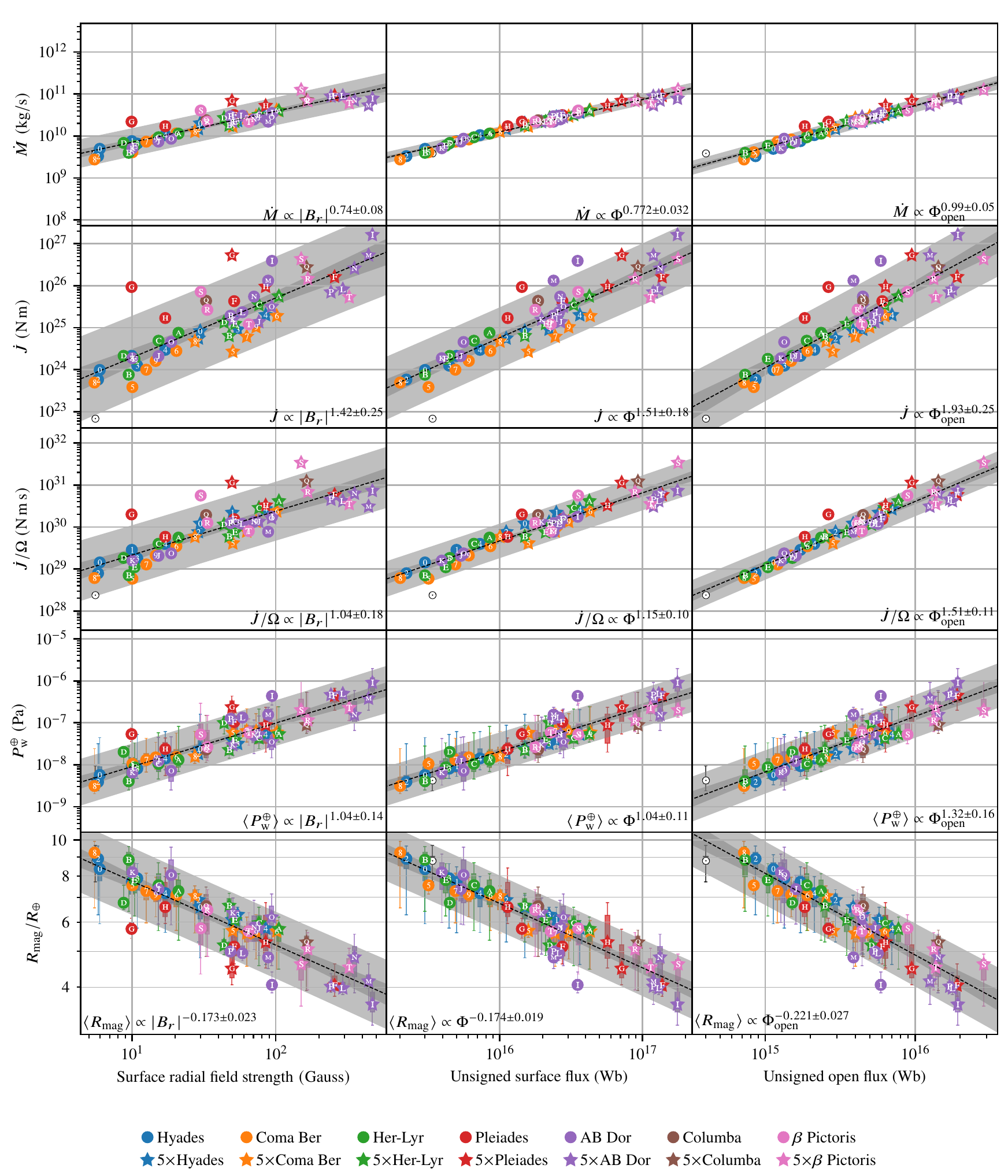}  %
    \caption{
        This figure is similar to Fig.~\ref{fig:trend-5x3-broken} except that the models of the \(B_\ZDI\) and \(5B_\ZDI\) series are treated as a single population of sixty stellar wind models. 
        Key statistical parameters of the fitted trend lines, variation bands, and prediction bands are given in Table~\ref{tab:correlations-pooled}.
    }\label{fig:trend-5x3} 
\end{figure*}
\begin{table*}
    \centering
    \caption{
        This table is similar to 
        Table~\ref{tab:correlations-split}, except that the models of the \(B_\ZDI\) and \(5B_\ZDI\) series are treated as a single population of sixty stellar wind models. The corresponding data and fitted parameters are plotted in Fig.~\ref{fig:trend-5x3}. 
    }\label{tab:correlations-pooled}
    \sisetup{
        table-figures-decimal=2,
        table-figures-integer=1,
        table-figures-uncertainty=2,
        table-number-alignment=center,
        add-decimal-zero=false,
        add-integer-zero = false,
        omit-uncertainty = false
    }
    \input{tables/correlations-rvalues-pvalues-predint-Full.tex}
\end{table*}
In \citet{paper1}, which considered five stars in the Hyades, the same analysis as in Section~\ref{sec:statistical-trends-and-correlations} was given for a single, pooled data series that comprises \(B_\ZDI\) series models and the \(5B_\ZDI\) series models.
Thus, in order to maintain consistency with \citet{paper1,paper2} we provide here the full pooled series analysis which includes the thirty \(B_\ZDI\) series models and the thirty \(5B_\ZDI\) series models, for a total of sixty wind models. The fitted trend lines and prediction bands are shown in Fig.~\ref{fig:trend-5x3} with the corresponding data in Table~\ref{tab:correlations-pooled}. 

For a large range of \(B_r\) values, such as what is presented here, it should be remembered that the radial magnetic field strength is correlated with the rotation period.

%% file: tables/correlations-rvalues-pvalues-predint-Full.tex
\begin{tabular}{
    l
    S[table-format=.2(2)]  %
    S[table-format=+1.2(2)]  %
    S[table-format=1.2]     %
    S[table-format=2.2]     %
    S[table-format=1.1e+2]   %
    }
    \toprule
{Quantity} & \multicolumn{5}{c}{Correlation with $a \log_{10}|B_r| + \log_{10} b$} \\

    \cmidrule(lr){2-6}
    {} & 
    {$a$} & {$b$} & {$r^2$} & {$\frac{y_\text{0.975}}{y_\text{0.025}}$} & {$p$} \\
    \midrule
$\log_{10}\max|B_r|$                                 &   1.06 \pm 0.06 &  0.51 \pm 0.10 &  0.96 &  2.81 &3.2e-42\\
$\log_{10}|\vec{B}|$                                 &   1.00 \pm 0.00 &  0.15 \pm 0.01 &0.9997 &  1.09 &2.2e-103\\
$\log_{10}\Phi$                                      &   0.99 \pm 0.08 & 14.67 \pm 0.14 &  0.91 &  4.21 &1.3e-32\\
$\log_{10}\Phi_\text{open}$                          &   0.74 \pm 0.08 & 14.39 \pm 0.13 &  0.86 &  4.17 &3.1e-26\\
$\log_{10} R_\Alfven$                                &   0.31 \pm 0.02 &  0.81 \pm 0.04 &  0.94 &  1.47 &2.3e-36\\
$\log_{10} |\vec{r_\Alfven} \times \uvec{\Omega}|$   &   0.30 \pm 0.02 &  0.71 \pm 0.04 &  0.93 &  1.49 &9.4e-35\\
$\log_{10}\dot M$                                    &   0.74 \pm 0.08 &  9.11 \pm 0.14 &  0.84 &  4.64 &9.9e-25\\
$\log_{10}\dot J$                                    &   1.42 \pm 0.25 & 22.87 \pm 0.43 &  0.69 & 94.74 &2.7e-16\\
$\log_{10}\dot J / \Omega$                           &   1.04 \pm 0.18 & 28.29 \pm 0.31 &  0.70 & 25.70 &8.2e-17\\
$\log_{10}P_\text{wind}^\Earth$                      &   1.04 \pm 0.14 & -9.08 \pm 0.24 &  0.79 & 12.27 &1.3e-21\\
$\log_{10} R_\text{mag}$                             &  -0.17 \pm 0.02 &  1.06 \pm 0.04 &  0.79 &  1.52 &1.3e-21\\
\midrule
{Quantity} & \multicolumn{5}{c}{Correlation with $a \log_{10}\Phi + \log_{10} b$} \\

    \cmidrule(lr){2-6}
    {} & 
    {$a$} & {$b$} & {$r^2$} & {$\frac{y_\text{0.975}}{y_\text{0.025}}$} & {$p$} \\
    \midrule
$\log_{10}|B_r|$                                     &   0.93 \pm 0.07 &-13.46 \pm 1.22 &  0.91 &  4.03 &1.3e-32\\
$\log_{10}\max|B_r|$                                 &   0.99 \pm 0.09 &-13.95 \pm 1.48 &  0.89 &  5.43 &8.5e-30\\
$\log_{10}|\vec{B}|$                                 &   0.93 \pm 0.07 &-13.36 \pm 1.22 &  0.91 &  4.05 &1.3e-32\\
$\log_{10}\Phi_\text{open}$                          &   0.76 \pm 0.03 &  3.23 \pm 0.57 &  0.97 &  1.92 &5.7e-46\\
$\log_{10} R_\Alfven$                                &   0.29 \pm 0.03 & -3.35 \pm 0.52 &  0.85 &  1.81 &2.4e-25\\
$\log_{10} |\vec{r_\Alfven} \times \uvec{\Omega}|$   &   0.28 \pm 0.03 & -3.34 \pm 0.51 &  0.85 &  1.80 &3.7e-25\\
$\log_{10}\dot M$                                    &   0.77 \pm 0.03 & -2.26 \pm 0.52 &  0.98 &  1.81 &9.5e-49\\
$\log_{10}\dot J$                                    &   1.51 \pm 0.18 &  0.57 \pm 2.94 &  0.83 & 28.90 &6.0e-24\\
$\log_{10}\dot J / \Omega$                           &   1.15 \pm 0.10 & 11.35 \pm 1.66 &  0.90 &  6.71 &2.6e-30\\
$\log_{10}P_\text{wind}^\Earth$                      &   1.04 \pm 0.11 &-24.39 \pm 1.83 &  0.86 &  8.11 &3.6e-26\\
$\log_{10} R_\text{mag}$                             &  -0.17 \pm 0.02 &  3.61 \pm 0.30 &  0.86 &  1.42 &3.6e-26\\
\midrule
{Quantity} & \multicolumn{5}{c}{Correlation with $a \log_{10}\Phi_\text{open} + \log_{10} b$} \\

    \cmidrule(lr){2-6}
    {} & 
    {$a$} & {$b$} & {$r^2$} & {$\frac{y_\text{0.975}}{y_\text{0.025}}$} & {$p$} \\
    \midrule
$\log_{10}|B_r|$                                     &   1.16 \pm 0.12 &-16.53 \pm 1.95 &  0.86 &  6.02 &3.1e-26\\
$\log_{10}\max|B_r|$                                 &   1.24 \pm 0.15 &-17.03 \pm 2.38 &  0.82 &  8.96 &3.0e-23\\
$\log_{10}|\vec{B}|$                                 &   1.17 \pm 0.13 &-16.41 \pm 1.97 &  0.86 &  6.15 &5.0e-26\\
$\log_{10}\Phi$                                      &   1.28 \pm 0.06 & -3.64 \pm 0.92 &  0.97 &  2.33 &5.7e-46\\
$\log_{10} R_\Alfven$                                &   0.37 \pm 0.04 & -4.48 \pm 0.64 &  0.85 &  1.81 &1.6e-25\\
$\log_{10} |\vec{r_\Alfven} \times \uvec{\Omega}|$   &   0.36 \pm 0.04 & -4.46 \pm 0.62 &  0.85 &  1.78 &1.2e-25\\
$\log_{10}\dot M$                                    &   0.99 \pm 0.05 & -5.20 \pm 0.80 &  0.96 &  2.09 &3.0e-43\\
$\log_{10}\dot J$                                    &   1.93 \pm 0.25 & -4.86 \pm 3.94 &  0.80 & 37.89 &5.5e-22\\
$\log_{10}\dot J / \Omega$                           &   1.51 \pm 0.11 &  6.48 \pm 1.76 &  0.92 &  5.09 &2.9e-34\\
$\log_{10}P_\text{wind}^\Earth$                      &   1.32 \pm 0.16 &-28.04 \pm 2.55 &  0.82 & 10.49 &3.1e-23\\
$\log_{10} R_\text{mag}$                             &  -0.22 \pm 0.03 &  4.22 \pm 0.42 &  0.82 &  1.48 &3.1e-23\\
\bottomrule
\end{tabular}